\documentclass[aps,onecolumn,nofootinbib,superscriptaddress,floatfix]{revtex4}

\usepackage{color,amsmath,amssymb,graphicx,latexsym,subfigure}

\usepackage{multirow}
\usepackage{ulem}
\usepackage{subcaption}

\usepackage{float}
\usepackage{booktabs}
\usepackage{threeparttable}
\usepackage[colorlinks,linkcolor=green, anchorcolor=green,citecolor=green]{hyperref}
\usepackage{ulem,soul}
\usepackage{xcolor}

\newcommand{\be}{\begin{equation}}
\newcommand{\ee}{\end{equation}}
\newcommand{\ba}{\begin{eqnarray}}
\newcommand{\ea}{\end{eqnarray}}

\newcommand{\gsim}{\mathrel{\hbox{\rlap{\lower.55ex \hbox {$\sim$}}
			\kern-.3em \raise.4ex \hbox{$>$}}}}
\newcommand{\lsim}{\mathrel{\hbox{\rlap{\lower.55ex \hbox {$\sim$}}
			\kern-.3em \raise.4ex \hbox{$<$}}}}

\makeatletter
\def\section{\@startsection {section}{1}{\z@}%
    {-3.5ex \@plus -1ex \@minus -.2ex}%
    {2.3ex \@plus.2ex}%
    {\normalfont\bfseries\boldmath\rightskip\z@}}
\makeatother

\hypersetup{colorlinks=true,
	breaklinks=true,
	pdfstartview=Fit,
	linkcolor=blue,
	citecolor=blue,
	urlcolor=blue}

\begin{document}

\title{A short review on Quintom dark energy theory \footnote{An invited review paper published in RAA}}

\author{Xin Ren} \email{rx@lzu.edu.cn}
\affiliation{Lanzhou Center for Theoretical Physics, Key Laboratory of Theoretical Physics of Gansu Province,
and Key Laboratory of Quantum Theory and Applications of MoE, Lanzhou University, Lanzhou 730000, China}
\affiliation{Institute of Theoretical Physics \& Research Center of Gravitation, School of Physical Science and Technology,
Lanzhou University, Lanzhou 730000, China}
\affiliation{Department of Astronomy, School of Physical Sciences,
 University of Science and Technology of China, Hefei 230026, China}

\author{Si-Yu Li}\email{lisy@ihep.ac.cn}
\affiliation{Key Laboratory of Particle Astrophysics, Institute of High Energy Physics, Chinese Academy of Sciences, Beijing 100049, China}

\author{Yang Liu}\email{liuy92@ihep.ac.cn}
\affiliation{Key Laboratory of Particle Astrophysics, Institute of High Energy Physics, Chinese Academy of Sciences, Beijing 100049, China}

\author{Yifu Cai} \email{yifucai@ustc.edu.cn}
\affiliation{Department of Astronomy, School of Physical Sciences,
 University of Science and Technology of China, Hefei 230026, China}
\affiliation{CAS Key Laboratory for Research in Galaxies and Cosmology,
 School of Astronomy and Space Science, University of Science and Technology of China, Hefei 230026, China}

\author{Hong Li}\email{hongli@ihep.ac.cn}
\affiliation{Key Laboratory of Particle Astrophysics, Institute of High Energy Physics, Chinese Academy of Sciences, Beijing 100049, China}
\affiliation{University of Chinese Academy of Sciences, Beijing 100049, People's Republic of China}

\author{Xinmin Zhang} \email{xmzhang@ihep.ac.cn}
\affiliation{Theoretical Physics Division, Institute of High Energy Physics, Chinese Academy of Sciences, 19B Yuquan Road, Shijingshan District, Beijing 100049, China}
\affiliation{School of Nuclear Science and Technology, University of Chinese Academy of Sciences, Beijing 101408, China}

\begin{abstract}
In this paper, we provide a short review on the Quintom dark energy theory. Firstly, we discuss the No-Go theorem associated with dynamical dark energy, then present some examples of models in which the equation of state (EoS) evolves with time and can cross $w=-1$ . Secondly, we discuss the bouncing universe and emergent universe with Quintom matter. Finally, we discuss the possibility of studying the nature of dark energy by measuring the Cosmic Microwave Background (CMB) polarization rotation angle. 
\end{abstract}
 
\maketitle

In 1998, the measurements of distances to high-redshift Type Ia supernovae (SN) led to the discovery of the accelerated expansion of the universe \cite{SupernovaSearchTeam:1998fmf, SupernovaCosmologyProject:1998vns}, a result that was further confirmed by the Cosmic Microwave Background (CMB) and other cosmological observations. Dark energy was introduced to describe this acceleration.

The equation of state (EoS) parameter of dark energy $w$ is a crucial parameter to characterize the properties of dark energy. In general, one can phenomenologically categorize dark energy models into the following main types according to the EoS parameter:
\begin{itemize}
  \item \textbf{Cosmological constant $\Lambda$}: The $w$ is fixed at $-1$ and the energy density of the dark energy remains constant over time.
  \item \textbf{Quintessence}: The $w$ lies above the cosmological constant boundary. The energy density of quintessence dark energy decays as the universe expands.
  \item \textbf{Phantom}: The $w$ lies below the cosmological constant boundary.  The energy density of the phantom dark energy would increase as the universe expands.
  \item \textbf{Quintom}: The $w$ can evolve across the cosmological-constant boundary $w=-1$, which is called Quintom behavior~\cite{Feng:2004ad} and is also referred to in the literature as phantom crossing or crossing the phantom divide~\cite{Hu:2004kh}. In the framework of Chevallier–Polarski–Linder (CPL) parameterization in Eq.~\eqref{eq:cpl}, Quintom dark energy can be classified into two types. \textbf{Quintom-A}  refers to the crossing from above (Quintessence) to below (Phantom) with time;  \textbf{Quintom-B} refers to the crossing from below (Phantom) to above (Quintessence) with time.
\end{itemize}

To constrain $w(a)$ with the cosmological data, it is usual to parameterize $w$ as a specific function of scale factor. One commonly used parameterization is the CPL parameterization \cite{Chevallier:2000qy, Linder:2002et}, which expresses $w$ as
\begin{equation}
 w(a)=w_0+w_a(1-a)~, \label{eq:cpl}
\end{equation}
or equivalently $w(z)=w_0+w_a z/(1+z)$.


\begin{figure}[htbp]
\begin{center}
\includegraphics[width=\textwidth]{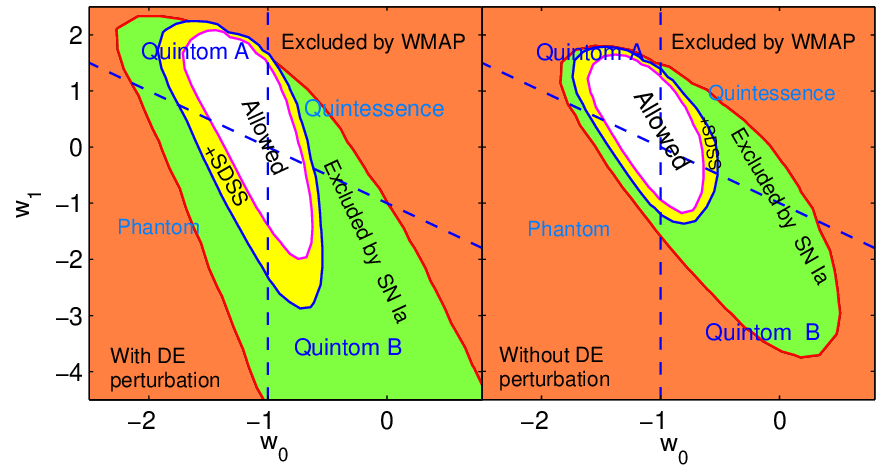}
\vskip-1.3cm \vspace{10mm}\caption{\it{Constraints in the $(w_0,w_1)$ plane from~\cite{Xia:2005ge}, where the original notation $w_1$ corresponds to $w_a$ in this review. 
    The left and right panels show the results with and without dark energy perturbations, respectively. 
    The dashed lines $w_0=-1$ and $w_0+w_a=-1$ separate the parameter space into the quintessence, phantom, Quintom-A and Quintom-B regions. 
    }
\label{fig:xia2006_w0wa}}
\end{center}
\end{figure}

In 2006, two of us, Zhang and Li with Xia, Zhao and Feng in~\cite{Xia:2005ge} provided a constraint on the $(w_0, w_a)$ parameters using the SN Ia, CMB, and large-scale structure data available at that time. (Note that $w_a$ is denoted by $w_1$ therein.) 
Fig.~\ref{fig:xia2006_w0wa} shows the resulting constraints in the $(w_0,w_a)$ plane. The left and right panels correspond to the cases with and without dark energy perturbations, respectively. The dashed lines $w_0=-1$ and $w_0+w_a=-1$ divide the parameter space into quintessence, phantom, Quintom-A, and Quintom-B regions. The comparison reveals that including dark energy perturbations visibly enlarges the allowed parameter region in the $(w_0,w_a)$ plane. This demonstrates that perturbations are not a negligible technical detail but can have a significant impact on the physical results. 

For a dynamical dark energy model whose EoS crosses the cosmological-constant boundary, the treatment of perturbations becomes a nontrivial issue. In the standard fluid description, the perturbation equations, Eqs.~\eqref{eq:delta perturbation} and~\eqref{eq:theta perturbation}, contain terms proportional to $(1+w)^{-1}$, leading to apparent singularities when $w=-1$ is crossed. Moreover, it was shown that conventional descriptions based on a single perfect fluid or a single field cannot realize a smooth Quintom crossing due to singularities or classical instabilities of perturbations around $w=-1$, which motivated the introduction of extra degrees of freedom in the Quintom theory \cite{Feng:2004ad}. 

To handle such crossings in practice, a parametrized treatment was proposed in \cite{Zhao:2005vj}. A very narrow interval around the crossing, $|1+w|<\epsilon$, is introduced, and the evolution is divided into three regions. In the outer regions, perturbations evolve with the standard equations on either side of the boundary; in the narrow intermediate region, the perturbations are assumed to remain finite and continuous and are matched by imposing $\dot{\delta}=0$ and $\dot{\theta}=0$. The width of this matching region is taken to be so small that neglecting the internal evolution provides a good approximation. The prescription was validated by comparing the resulting CMB and LSS power spectra with those from a full two-field Quintom model, finding negligible differences. This confirmed that perturbations in crossing models can be treated reliably within such a parametrized approach. The approach was later revisited in Ref.~\cite{Li:2010hm} and was shown to be well justified. Subsequently, another approach, called Parameterized Post-Friedmann (PPF), was also proposed for studying this kind of dark energy perturbations~\cite{Fang:2008sn}. 

\begin{figure*}[htbp]
    \centering
        \includegraphics[width=\textwidth]{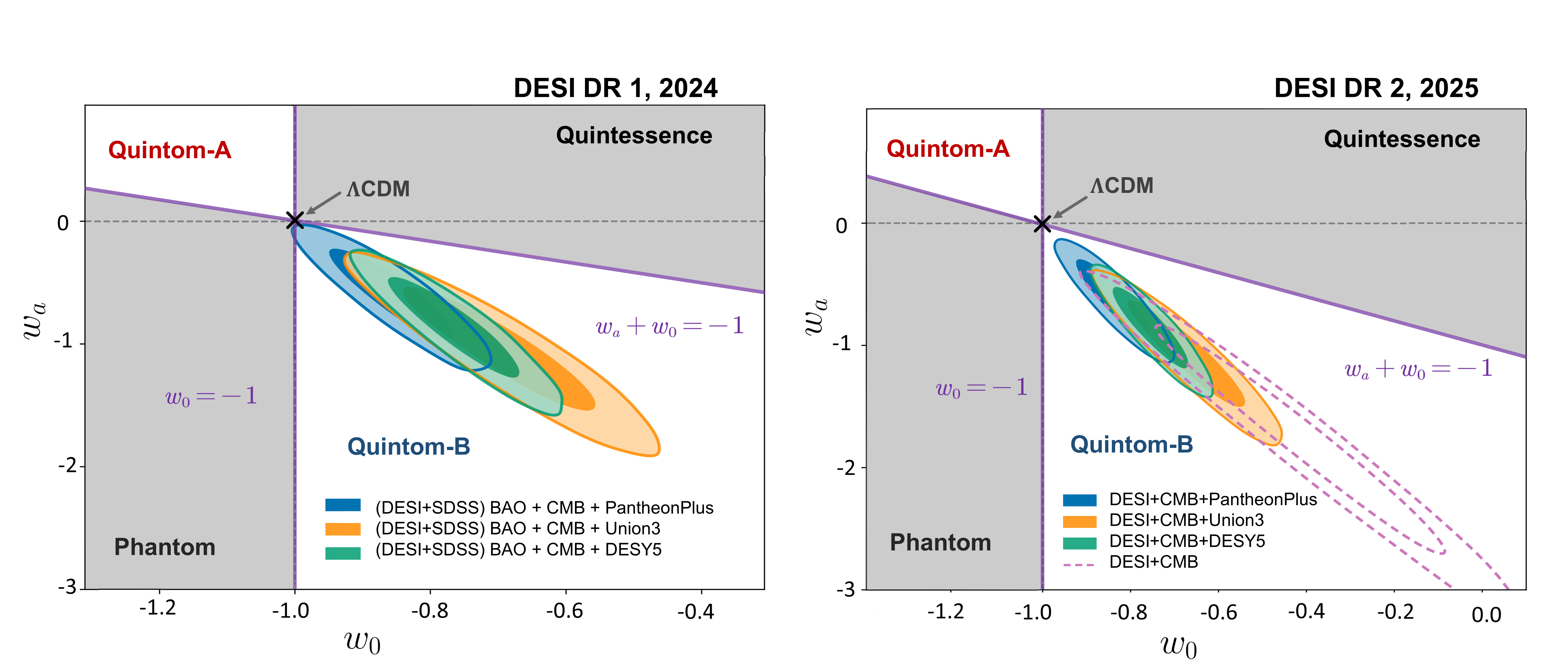}
    \caption{\it{The posterior constraints on the CPL parameterization $(w_0,w_a)$ of dynamical dark energy derived from the combination of DESI BAO, CMB, and three SN datasets. The left panel is taken from Ref.~\cite{DESI:2024mwx}, while the right panel is from Ref.~\cite{DESI:2025zgx}. {The two boundary lines $w_0=-1$ and $w_0+w_a=-1$ divide the parameter space into four categories.} And the 95\% contour plots for the two datasets show that the allowed parameter ranges fall mainly within the Quintom-B region.}}
    \label{fig:desi_w0wa_all}
\end{figure*}

In recent years, the Dark Energy Spectroscopic Instrument (DESI) team has used the most precise measurement data to create the largest three-dimensional map of the universe to date. They have measured the late-time expansion history of our universe with an accuracy of better than 1\%, providing us with a powerful approach to study the evolution of our universe. In April 2024, DESI released the results of its first year of observations,  suggesting dynamical dark energy with $2.5\sigma$, $3.5\sigma$, $3.9\sigma$ confidence when combined with CMB and each of {PantheonPlus}, Union3, or DESY5 SN datasets respectively~\cite{DESI:2024mwx}. This result has attracted considerable attention and extensive discussion~\cite{Cortes:2024lgw, DESI:2024aqx, DESI:2024kob,Wang:2024dka, Giare:2024gpk, Shlivko:2024llw, Ye:2024ywg, Tada:2024znt, Carloni:2024zpl, Gialamas:2024lyw, Luongo:2024fww, Giare:2024smz, Mukherjee:2024ryz, Jiang:2024xnu, Dinda:2024kjf, Yang:2024kdo, Yang:2025kgc, Giare:2024oil, Liu:2024gfy, Wang:2024hks, Park:2024vrw, Bhattacharya:2024hep, Thompson:2024nxf, Reboucas:2024smm, Andriot:2024jsh, Orchard:2024bve, Pang:2024qyh, Colgain:2024xqj, Colgain:2024mtg, Ishak:2024jhs, Zheng:2024qzi, Escamilla-Rivera:2024sae, Yin:2024hba, Chudaykin:2024gol, Huang:2025som,RoyChoudhury:2024wri,Wolf:2024stt,Wolf:2025jlc,Li:2025vuh,Li:2025dwz,Du:2025xes,Du:2026cly,Mishra:2026tzn}. The analysis of DESI Data Release~2, combined with CMB and supernova constraints, increases the statistical significance of the deviation from a cosmological constant to $2.8\sigma$, $3.8\sigma$ and $4.2\sigma$ 
for {PantheonPlus}, Union3 and DESY5 SN samples, respectively~\cite{DESI:2025zgx, DESI:2025fii,DESI:2025wyn}, providing support for Quintom-B scenario.

Fig.~\ref{fig:desi_w0wa_all} summarizes the constraints on the $(w_0, w_a)$ from the two DESI data releases.  In the same manner as Fig.~\ref{fig:xia2006_w0wa}, two boundary lines are introduced to divide the parameter space into four dark energy categories:  quintessence, phantom, Quintom-A, and Quintom-B. The intersection of the two dividing lines represents the $\Lambda$CDM  model. Notably, the observational results show support for Quintom-B.

\begin{figure*}[htbp]
    \centering
        \includegraphics[width=0.8\textwidth]{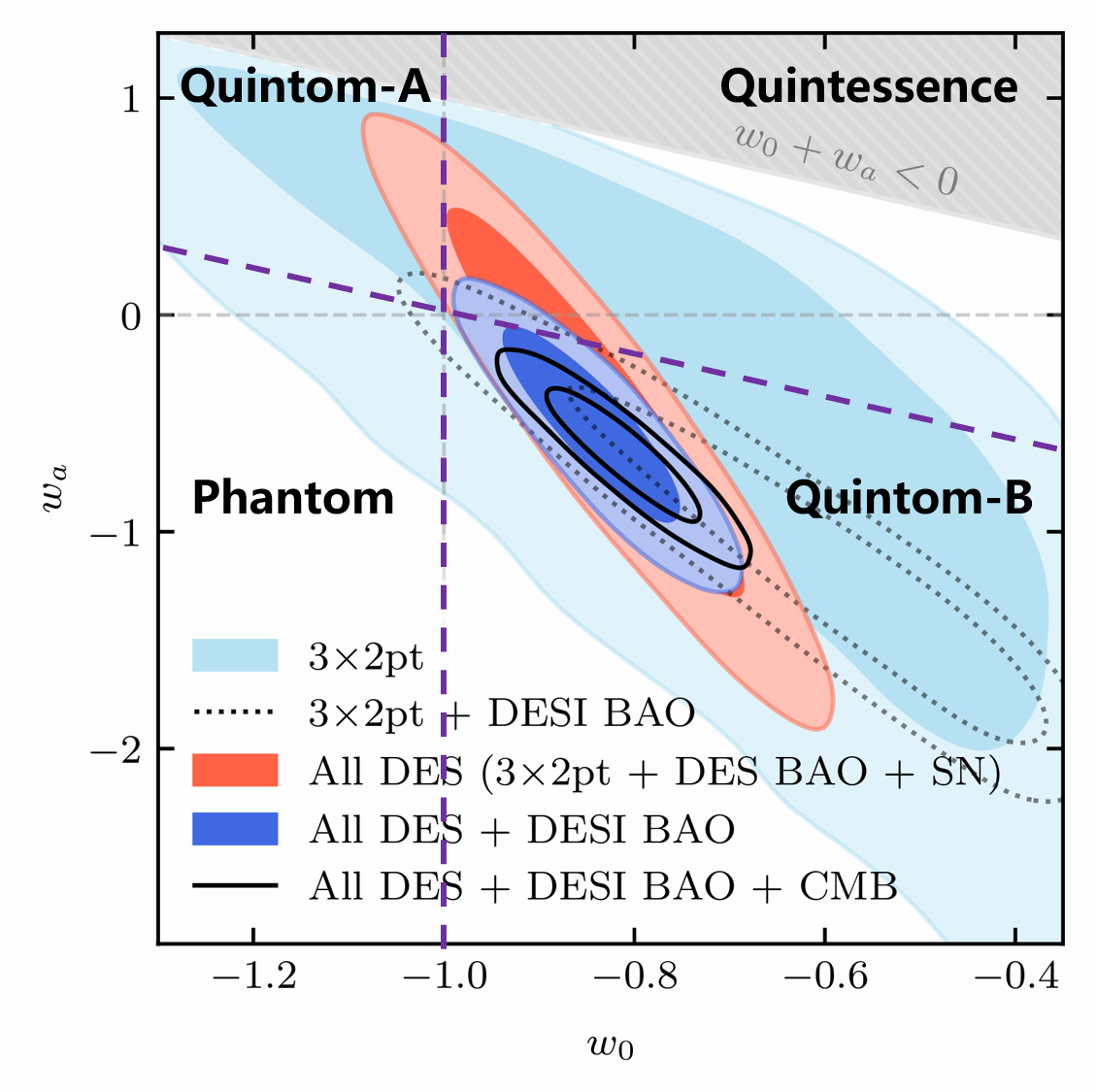}
    \caption{\it{The posterior constraints on the $(w_0,w_a)$ parameterization  derived from multiple dataset combinations with DES (3×2pt +DES BAO +SN), DESI BAO and CMB. The figure is taken from Ref.~\cite{DES:2026jmi}. {The boundary lines $w_0=-1$ and $w_0+w_a=-1$ are added to show the regions of quintessence, phantom, Quintom-A and Quintom-B.} }}
    \label{fig:des}
\end{figure*}

Very recently, the Dark Energy Survey (DES) collaboration has reported constraints on dynamical dark energy using its full six-year data set, combining type Ia supernovae, baryon acoustic oscillations, CMB, and weak gravitational lensing and galaxy clustering ($3\times 2$pt)~\cite{DES:2026jmi}.
As shown in Fig.~\ref{fig:des}, the constraint results from all datasets used by the DES collaboration also lie in the Quintom-B region, which is consistent with the result from DESI and favors Quintom-B.

The CPL parameterization essentially cuts off the Taylor expansion beyond the first order, making it a purely low-redshift description that imposes no constraints from higher-order terms. This could introduce a bias when higher-order effects are non-negligible. In other words,  parameterization methods always introduce certain assumptions, making the resulting constraints, to some extent, model-dependent.

\begin{figure*}[htbp]
    \centering
    \begin{minipage}{0.45\textwidth}
        \includegraphics[width=\textwidth]{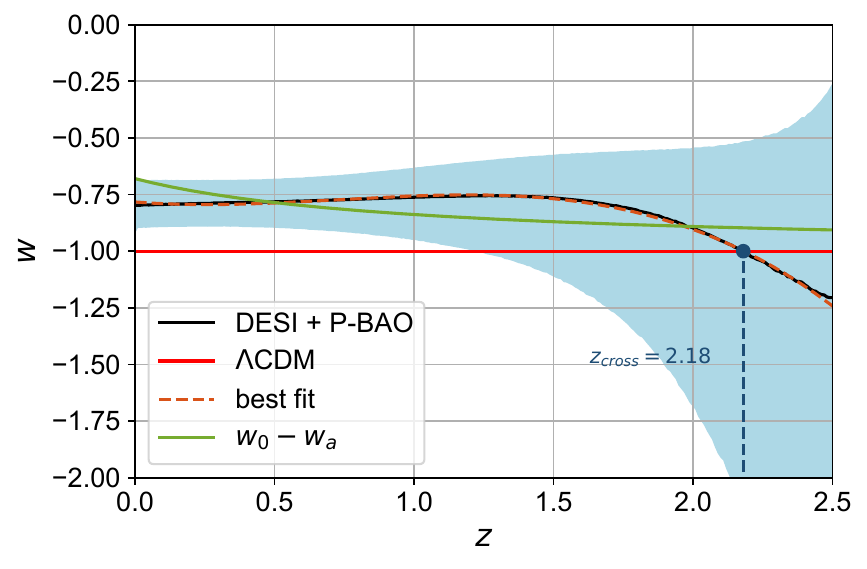}
    \end{minipage}
    \hfill
    \begin{minipage}{0.5\textwidth}
        \includegraphics[width=\textwidth]{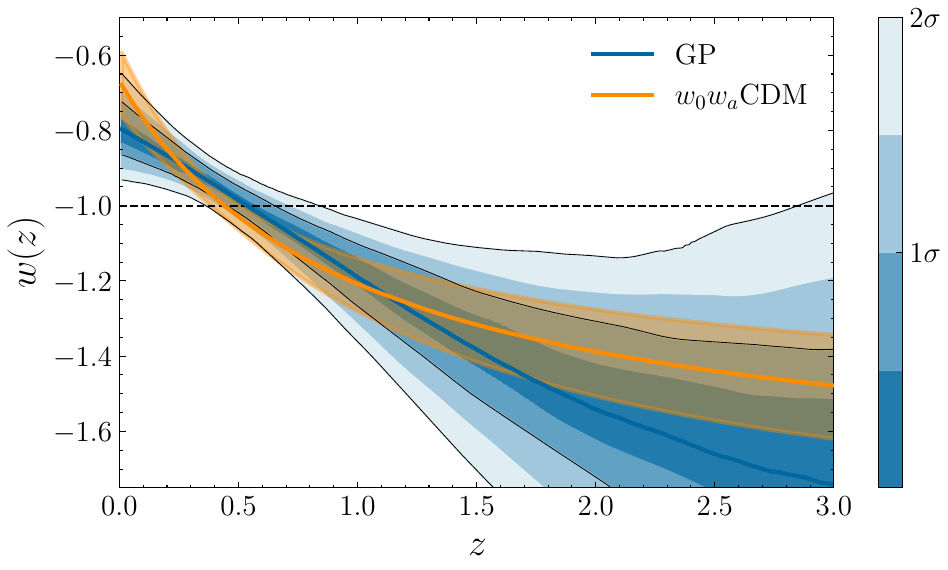}
    \end{minipage}
    \caption{\it{ The left panel is from Fig.2 of~\cite{Yang:2024kdo}, which shows the reconstructed EoS parameter $w$ for DESI DR1 BAO data and previous BAO data using Gaussian process regression. The right panel is from Fig.10 of \cite{DESI:2025fii}, showing the Gaussian process reconstruction and CPL parameterization result of EoS utilizing data from the combination of DESI, CMB and Union3.}}
    \label{fig:desi_gp}
\end{figure*}

Thus, non-parametric approaches should be considered. One widely used non-parametric method is the Gaussian process regression \cite{Shafieloo:2012ht, Holsclaw:2010sk, Holsclaw:2010nb, Seikel:2012uu}. Gaussian process regression provides a data-driven approach to reconstruct cosmological functions and their derivatives directly from observations without assuming a specific model. The reconstruction result is controlled by the covariance kernel, whose hyperparameters are determined by the observational data. In 2024, two of us, Cai and Ren together with Yang, Wang, Lu, Zhang and Saridakis used this method to reconstruct the evolution history of the Hubble parameter by combining the BAO data from DESI DR1 with previous BAO data, and then obtained a model-independent reconstruction of the dark energy EoS parameter~\cite{Yang:2024kdo}. In 2025, the DESI collaboration presented the Gaussian process reconstruction of $w(z)$ with the combined dataset of DESI DR2 BAO, CMB and supernovae~\cite{DESI:2025fii}. Fig.~\ref{fig:desi_gp} presents the reconstructed $w(z)$ results obtained from the Gaussian process regression method. These Gaussian process reconstruction results, along with the CPL parameterization, consistently show the Quintom-B behavior.

\begin{figure*}[htbp]
    \centering
        \includegraphics[width=\textwidth]{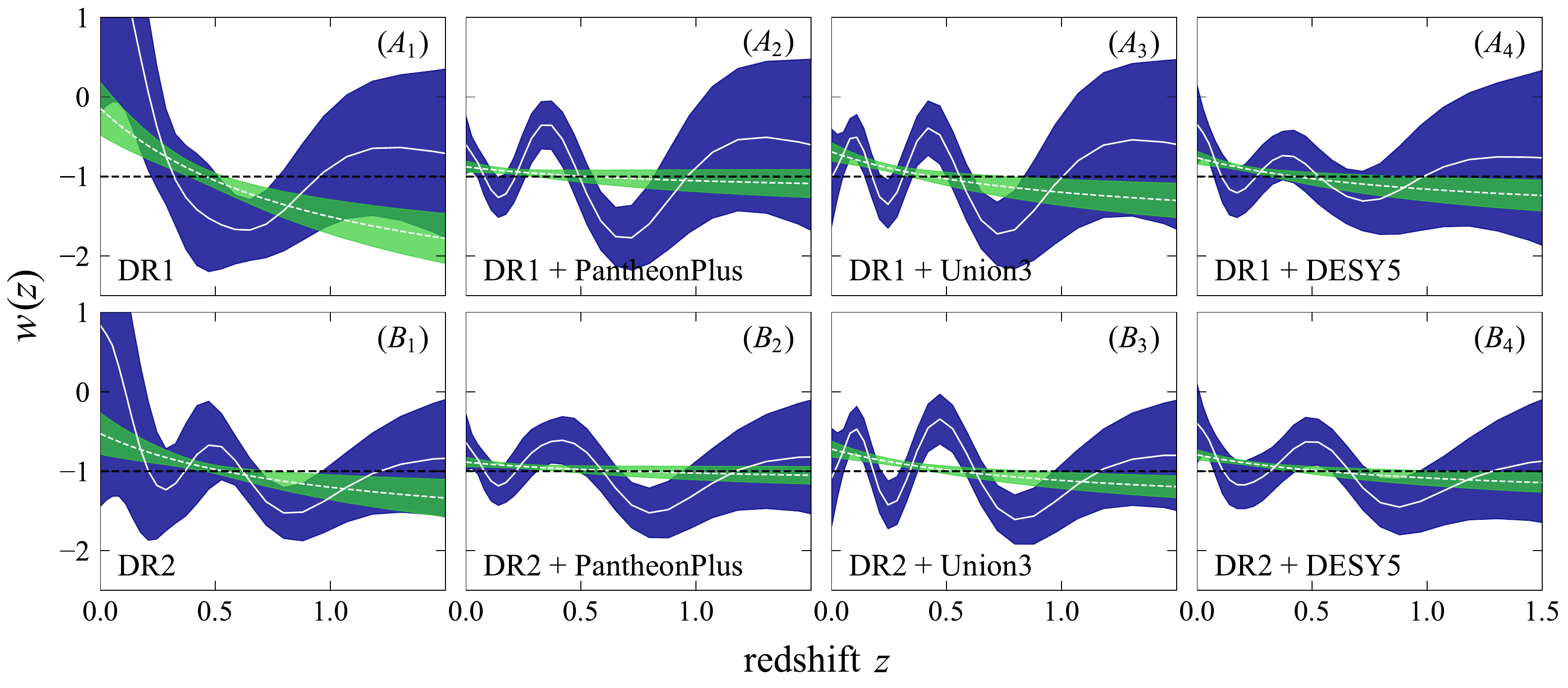}
    \caption{\it{{Dark energy EoS $w(z)$ reconstructed from several datasets-DESI DR1/DR2 BAO data, alone and combined with PantheonPlus, Union3, and DESY5 SN samples.}  The results are for two different approaches: the correlation-prior method (bottom-layered, dark-blue band) and the ($w_0$, $w_a$) parameterization (top-layered, green band). The panel is from Fig.3 of~\cite{DESI:2025wyn}. }}
    \label{fig:na_w}
\end{figure*}

Principal Component Analysis~\cite{Huterer:2002hy} is a transformation that provides a new basis in which the coefficients  corresponding to the bin amplitude parameters are decorrelated, typically achieved by diagonalizing the associated covariance matrix to identify uncorrelated eigenmodes.
In 2007 Gong-Bo Zhao, Dragan Huterer and Xinmin Zhang studied the dynamical features of dark energy through this Principal Component Analysis method~\cite{Zhao:2007ew}, and found previous data at that time showed  no significant evidence for dynamical dark energy. A non-parametric Bayesian method that applies a correlated prior for $w(z)$ reconstructed the evolution  of dark energy EoS in 2012~\cite{Zhao:2012aw}. 
This analysis revealed a preference for dynamical dark energy at the $2.5 \sigma$ level using combined data with SNLS3 and weak prior. The statistical significance of the deviation from the $\Lambda$CDM model rose to $3.5 \sigma$ in 2017, driven by the high signal-to-noise ratio and the expanded redshift coverage of the ALL16 dataset. Principal Component Analysis played a crucial role in this finding by isolating the data-driven eigenmodes of dark energy EoS~\cite{Zhao:2017cud}.
Most recently, the DESI collaboration employed shape-function methods and non-parametric Bayesian reconstruction with Principal Component Analysis to probe the dark energy EoS.
Their findings reveal an oscillatory around $w=-1$, with statistical evidence for dynamical dark energy exceeding a significance level of $4.3 \sigma$ for DESI DR2 BAO + DESY5, $3.9 \sigma$ for DESI DR2 BAO + Union3 and $3.1 \sigma$ for DESI DR2 BAO + PantheonPlus~\cite{DESI:2025wyn} shown in Fig.~\ref{fig:na_w}. These results, yielding conclusions consistent with other DESI publications, suggest that dark energy may be time-dependent and EoS cross the $w=-1$, providing robust constraints on the evolution of dark energy. 

The first dark energy model describing the EoS crossing $-1$ was put forward in April 2004 dubbed Quintom by the team of Xinmin Zhang \cite{Feng:2004ad}. That paper states "in general to realize the transition of $w$ around -1, one needs to consider models of dark energy with more
complicated dynamics and interactions with gravity and
matter. This class model of dark energy, which we dub
“Quintom”, is different from the quintessence or phantom in the determination of the evolution and fate of the universe".  In June 2004, Xinmin Zhang was invited to give a plenary talk on Quintom dark energy in the "SUSY2004" held in Tsukuba, Japan~\cite{Zhang:2004gb}. In October 2004, Wayne Hu wrote a paper on the behavior of the dark energy EoS crossing -1 based on a parametrized $w(z)$, referring to it as 'crossing the phantom divide'\cite{Hu:2004kh}.

Before discussing in detail the realizations of Quintom dark energy models, we revisit the No-Go theorem associated with dynamical dark energy. {The No-Go theorem states that, in the context of basic single perfect fluid models or single-field models with a Lagrangian $L = L(\phi, \partial_\mu \phi\partial^\mu\phi)$, which means the scalar field have single degree of freedom without higher derivatives, minimally coupled to general relativity in the Friedmann-Robertson-Walker universe, the equation of state parameter 
$w$ cannot cross the cosmological constant boundary $-1$~\cite{Feng:2004ad,Vikman:2004dc,Zhao:2005vj,Cai:2009zp}. This can be shown through the stability analysis of dark energy perturbation, in which one typically assumes a perfect fluid description for calculations.}
In conformal Newtonian gauge $ds^2=-a(\eta)^2 [ -(1+2\Psi) {\rm d} \eta^2+(1-2\Phi)\delta_{ij} {\rm d} x^i {\rm d} x^j ]$ assuming no anisotropic perturbations, which leads to $\Psi=\Phi$. The description of dark energy perturbations in Fourier space is given by~\cite{Kodama:1984ziu,Ma:1995ey}
\begin{align}
    \delta '&=- \Big (1+\frac{\bar{p}}{\bar{\rho}}\Big)(\theta-3\Phi')-3\mathcal{H} \Big (\frac{\delta p}{ \delta \rho}-\frac{\bar{p}}{\bar{\rho}} \Big )\delta~, \label{eq:delta perturbation} \\
    \theta'&=-\Big(\mathcal{H}+\frac{\bar{p}'}{\bar{\rho}+\bar{p}}\Big)\theta+ k^2 \Big ( \frac{\delta p}{\bar{\rho}+\bar{p}} + \Psi \Big)~,  \label{eq:theta perturbation}
\end{align}
with $ad\eta=dt$ defining the conformal time $\eta$ and denoting derivatives with respect to it by a prime. $\mathcal{H}$ is the conformal Hubble parameter, while the density contrast and the velocity perturbation are $\delta \equiv \delta \rho / \bar{\rho}$, $\theta \equiv ik^{j}\delta T^{0}_{j}/(\bar{\rho}+\bar{p})$, respectively. 

For a barotropic perfect fluid, its adiabatic sound speed is given by:
\begin{equation}
    c_a^2=\frac{\delta p}{\delta \rho}\Big |_{\rm{adiabatic}}=\frac{\bar{p}'}{\bar{\rho}'}=w-\frac{w'}{3\mathcal{H}(1+w)}.
\end{equation}
with EoS parameter defined as $w=\bar{p}/\bar{\rho}$. In this case, a divergence in the sound speed expression occurs when $w$ crosses $-1$.
{This divergence signals a breakdown of the fluid description and leads to unphysical instability in the evolution of dark energy perturbations.}

For the case of a non-barotropic fluid, the entropy variation naturally arises. Consequently, dark energy perturbations generate non-adiabatic modes, making the simple expression of the sound speed no longer adequate.  Instead, a more general definition based on the relation between the gauge invariant parameters becomes necessary.

Thus we could define a more general relationship between the pressure and the energy
density
\begin{equation}
    c_s^2=\frac{\delta \hat{p}}{\delta \hat{\rho}}.
\end{equation}
with gauge invariant perturbation of pressure $\delta \hat{p}$ and  density fluctuation $\delta \hat{\rho}$
\begin{align}
    \delta \hat{p}&=\delta p+3\mathcal{H}c_a^2(1+w)\bar{\rho}\frac{\theta}{k^2}, \\
    \delta \hat{\rho}&=\delta \rho+3\mathcal{H}(1+w)\bar{\rho}\frac{\theta}{k^2}.
\end{align}
Consequently, we derive the relation between $\delta p$ and $\delta \rho$ in a general frame as
\begin{equation}
    \delta p = c_s^2 \delta \rho +(c_s^2-c_a^2) \Big[ 3 \mathcal{H}(1+w)\bar{\rho} \Big]\frac{\theta}{k^2}.
\end{equation}
Substituting this relation into Eq.~\eqref{eq:theta perturbation}, we get
\begin{equation}
\begin{aligned}
\theta'&=-\Big [\mathcal{H}-3\mathcal{H}(c_s^2-c_a^2+w)+\frac{w'}{1+w} \Big ]\theta\\ &+k^2 \Big (\frac{c_s^2}{1+w}\delta + \Psi \Big) \\
 &=-\mathcal{H}(1-3w)\theta+k^2 \Psi\\ &+\frac{1}{1+w}\Big[3\mathcal{H}(1+w)(c_s^2-c_a^2)\theta-w'\theta+k^2 c_s^2\delta \Big ] \\
 &=-\mathcal{H}\theta+k^2\Psi+\frac{k^2\delta \hat{p}}{(1+w)\bar{\rho}}.
\end{aligned}
\label{eq:non-ad v-perturbation}
\end{equation}

The gauge invariant intrinsic entropy perturbation $\hat{\Gamma}$ takes the form as
\begin{equation}
    \hat{\Gamma}=\frac{1}{w\bar{\rho}}(\delta p -c_a^2 \delta \rho)=\frac{1}{w \bar{\rho}}(\delta \hat{p}-c_a^2\delta \hat{\rho}).
\end{equation}

When crossing the cosmological constant boundary, the velocity perturbations $\theta$ and $\theta'$ will be divergent, unless we impose $\delta \hat{p}=0$ at that point.

In order to keep $\hat{\Gamma}$ finite, the divergence of the adiabatic sound speed $c_a^2$ forces us to take $\delta \hat{\rho}=0$ when crossing the $-1$ boundary. This will give the relationship of $\delta p = c_a^2\delta \rho$, which is incompatible with the non-adiabatic perturbations we originally assumed. Therefore, the perturbation system becomes ill-defined around the crossing point, indicating that a regular transition across $w = -1$ cannot be consistently realized within the framework of non-adiabatic perturbations.
{In summary, a single perfect fluid is insufficient to achieve the crossing of $w=-1$ from classical stability analysis.} The same conclusion applies as well to the case of a generic single scalar field without higher derivatives.

Due to the No-Go theorem, one needs to introduce additional degrees of freedom such as multiple fields, higher derivative terms, extended theories of gravity and interacting dark energy to realize the crossing of $w=-1$.

The two-field Quintom model is a combination of a quintessence field and a phantom field, and its action can be expressed as:
\begin{equation}
 S=\int d^4 x \sqrt{-g}\left[-\frac{1}{2} \nabla_\mu \phi \nabla^\mu \phi+\frac{1}{2} \nabla_\mu \sigma \nabla^\mu \sigma-V(\phi, \sigma)\right].
\end{equation}
Thus, the effective energy density $\rho$ and the effective pressure $p$ are given by
\begin{equation}
\rho=\frac{1}{2} \dot{\phi}^2-\frac{1}{2} \dot{\sigma}^2+V(\phi, \sigma),\quad  p=\frac{1}{2} \dot{\phi}^2-\frac{1}{2} \dot{\sigma}^2-V(\phi, \sigma),
\end{equation}
and the corresponding EoS is now given by
\begin{equation}
w=\frac{p}{\rho}=\frac{\dot{\phi}^2-\dot{\sigma}^2-2 V(\phi, \sigma)}{\dot{\phi}^2-\dot{\sigma}^2+2 V(\phi, \sigma)}.
\end{equation}

In the original two-field model, the potential was chosen to take an exponential form without direct coupling $V(\phi, \sigma)=V_\phi(\phi)+V_\sigma(\sigma)=V_{\phi 0} e^{-\lambda_\phi \kappa \phi}+V_{\sigma 0} e^{-\lambda_\sigma \kappa \sigma}$~\cite{Feng:2004ad,Guo:2004fq}. The two-scalar field quintom model with Gaussian potential and hyperbolic tangent potential is also used to explain the DESI DR2 data, achieving the Quintom-B behavior favored by observations~\cite{Goh:2025upc}. In addition to these, numerous related models have been further considered and studied, including two-scalar field models with other different potential forms~\cite{Perivolaropoulos:2004yr,Zhang:2005eg,Zhao:2006mp, Chimento:2008ws} or mixed kinetic terms~\cite{Saridakis:2009jq}, complex scalar fields~\cite{Wei:2005nw,Wei:2005fq,Wei:2007rp} and multiple scalar fields~\cite{Mughal:2020glg,Vazquez:2023kyx}.

In addition to introducing new degrees of freedom by adding more fields,  the inclusion of higher derivative operators can also bring in new degrees of freedom and provide an alternative way to realize the Quintom scenario within a single-field framework~\cite{Li:2005fm, Zhang:2006ck, Cai:2007gs}.

The Lagrangian of a single field model with  higher derivatives can be expressed in the following form
\begin{equation}
\mathcal{L}=\mathcal{L}\left(\phi, X, \square \phi \square \phi, \nabla_\mu \nabla_\nu \phi \nabla^\mu \nabla^\nu \phi\right).
\end{equation}

The more general scalar-tensor theories, including Galileon~\cite{Nicolis:2008in, Deffayet:2009wt, Deffayet:2011gz}, Horndeski~\cite{Horndeski:1974wa} and their extension DHOST~\cite{Langlois:2017mxy, Langlois:2018jdg}, can also realize the Quintom scenario for dark energy. The actions of these theories involve couplings between the scalar field, its kinetic term, and higher-order derivative terms.  This allows the EoS parameter to exhibit more complex evolution, realizing a smooth crossing of $w=-1$ while avoiding ghost instabilities \cite{Li:2011qfa,Matsumoto:2017qil, Tiwari:2024gzo}.

Alternatively, beyond introducing extra matter fields, the Quintom scenario can also be realized through the modified gravity theory. In this approach, the effective dark energy behavior emerges from the extended geometric part of the action. Under appropriate function forms, the effective EoS can naturally cross the $-1$. Early realizations of Quintom behavior within modified gravity were achieved in frameworks such as Gauss-Bonnet gravity~\cite{Cai:2005ie,Nojiri:2005jg,Leith:2007bu} and $f(R)$ gravity~\cite{Starobinsky:1980te, Capozziello:2002rd,Nojiri:2010wj}. This has later been generalized to metric teleparallel theories including $f(T)$ gravity~\cite{Cai:2015emx, Bahamonde:2021gfp, Krssak:2018ywd,Wu:2010av,Ren:2021tfi,Briffa:2020qli,Yang:2025mws} and symmetric teleparallel theories including $f(Q)$ gravity~\cite{ BeltranJimenez:2019tme, Heisenberg:2023lru, Koussour:2023ulc,Yang:2025kgc,Basilakos:2025olm}.

The interacting dark energy model provides another viable approach to realize the Quintom scenario. It describes the existence of interactions between the dark energy and matter sectors. Energy transfer between the two sectors can effectively modify the EoS of dark energy. With suitable forms of interaction term $Q_{\mathrm{int}}$, the effective EoS can exhibit Quintom-like behavior. Interacting dark energy models have received extensive study~\cite{Amendola:1999er,Zhang:2004gb,Cai:2004dk,Das:2005yj,Zhang:2005eg, Zhao:2009ke,Wang:2005jx, Wang:2005ph, Yang:2018uae, Linder:2025zxb,Zhang:2005kj}, and after DESI, this dynamical dark energy model capable of realizing the crossing of $w=-1$ has also drawn considerable attention~\cite{Li:2024qso,Pan:2025qwy,Silva:2025hxw,Li:2025ula,Chakraborty:2025syu,Li:2025owk, Khoury:2025txd, Wang:2025znm, Li:2026xaz}.

The realization of the Quintom scenario can also be associated with other fundamental physical theories, such as some string-inspired models~\cite{Cai:2007gs,Sadeghi:2008qp,Zhang:2008ac} and models related to axion fields~\cite{Khoury:2025txd,Mishra:2025goj,Toomey:2025yuy}. There are also studies exploring the realization of the Quintom scenario via fermion fields~\cite{Cai:2008gk, Wang:2009ae, Wang:2009ag, Dil:2016vod}, interacting holographic Quintom dark
energy~\cite{Li:2024qus, Zhang:2005yz, Zhang:2006qu} and higher-dimensional gauge field~\cite{Koutroulis:2026gjr}. These relevant theories that can realize the Quintom scenario have also been introduced in previous related review articles~\cite{Cai:2009zp,Cai:2025mas}.

The Quintom scenario can not only describe the evolutionary characteristics of dark energy in the late universe but can also be applied to the early universe. In 2007, Quintom matter was introduced into the very early universe to realize non-singular bounce models~\cite{Cai:2007qw}. In these models, the universe initially undergoes a contracting phase and then evolves into an expanding phase. The point connecting these two phases is called the bounce point, which corresponds to a finite minimum scale factor, thereby avoiding the singularity present in the Big Bang model.

In a non-singular bounce cosmology, the universe goes from a contracting phase with $\dot{a}<0$ and $H<0$ to an expanding phase with $\dot{a}>0$ and $H>0$, passing through a bounce point where $H=0$. From the Friedmann equation $\dot{H} = -4\pi G(\rho+p)$, one can see that a successful bounce requires $\dot{H}>0$ and thus $\rho+p<0$, namely $w<-1$ at the bounce. In the early contracting stage, however, the universe is dominated by matter or radiation, giving $w>-1$. Therefore, in order to reach the phantom condition $w<-1$, the total equation of state must first cross $w=-1$ from above. After the bounce, the universe enters an expanding phase and must eventually connect to the standard radiation-dominated ($w=1/3$) and matter-dominated ($w=0$) eras. This means that the equation of state has to cross $w=-1$ a second time, now from below, returning to the normal region $w>-1$. As a result, in a complete non-singular bounce the universe's total EoS crosses $w=-1$ twice, once downward and once upward. Such a crossing behavior is the characteristic property of the Quintom model, which combines both quintessence and phantom components.

The Quintom bounce models can be realized through two-field models, higher derivative models, modified gravity, and other frameworks~\cite{ Cai:2008qb, Cai:2008qw, Setare:2008qr, Zhang:2007bi, Cai:2011tc, Cai:2012va, Alexander:2014uaa,Kolevatov:2017kvx, Singh:2022jue,Tukhashvili:2023itb}. The corresponding perturbation behaviors have also been studied~\cite{Cai:2007zv, Cai:2008ed,  Cai:2009fn, Karouby:2011wj}. Further details can be found in relevant review articles~\cite{Qiu:2025oop}.

Based on the bounce scenario, a cyclic universe model can be further constructed. A cyclic universe can be viewed as a combination of multiple bounce models: after undergoing one bounce, the universe enters an expanding phase, from which it may later return to a contracting phase and repeat the previous bouncing fate, thereby entering a cyclic evolutionary state~\cite{Steinhardt:2002ih, Brown:2004cs, Piao:2004me, Zhang:2007an, Xiong:2007cn, Itzhaki:2025gdv}. In this scenario, the EoS parameter $w$ of the overall cosmic components exhibits repeated crossings of $w = -1$, displaying repetitive Quintom-like evolutionary behavior, and thereby avoids potential singularities in cosmic evolution.

Another model that can circumvent the singularity present in the Big Bang model is known as the emergent universe model~\cite{Ellis:2002we,Ellis:2003qz, Cai:2012yf, Cai:2013rna, Ilyas:2020zcb}. It describes a universe that initially stays in a static and finite phase , where both the Hubble parameter $H$ and the overall energy density of the universe tend to zero. Subsequently, the universe enters an expanding phase. Similarly to the transition from the bounce point to the expansion phase in bounce models, the emergent universe also requires the cosmic components to exhibit Quintom-like behavior in order to connect to the subsequent conventional epoch of cosmic expansion.

These early universe models suggest that the Quintom scenario is not limited to describing dark energy and has broader applicability in cosmology.

In addition to studying dark energy through its gravitational effects on the cosmic expansion history and structure formation, one may also explore its possible interactions with ordinary matter. If dark energy is described by a dynamical scalar field $\phi$, its interaction with Standard Model particles can be parametrized in an effective field theory form. Imposing a shift symmetry $\phi\rightarrow \phi+{\rm const.}$ to evade the experimental constraints on fifth-forces, the leading-order interaction can be written as
\begin{equation}
    {\cal L}_{\rm int}=\sum_i c_i \partial_\mu \phi J_i^\mu ,
\end{equation}
where $J_i^\mu$ denotes a current associated with the Standard Model particles.

Taking the current $J_i^\mu$ to be the electromagnetic Chern--Simons current, the interaction in the above equation becomes
\begin{equation}
    {\cal L}_{\rm CS}\sim \partial_\mu\phi A_\nu \tilde{F}^{\mu\nu},
\end{equation}

Such an interaction leads to a rotation of the polarization direction of propagating photons, giving rise to the phenomenon commonly referred to as cosmic birefringence. In this framework, the observed rotation angle is determined by the difference in the scalar-field value between the emission and observation epochs~,
\begin{equation}
    \beta \propto \phi_0-\phi_{\rm em},
\end{equation}
where $\phi_{\rm em}$ and $\phi_0$ are the values of the dark energy scalar at photon emission and today, respectively. Therefore, the measurement of the polarization rotation angle provides a way to probe the dynamics of the dark energy scalar field and its Chern--Simons interaction with photons.

The CMB is the oldest polarized light in the Universe, and hence provides a particularly useful probe of this effect. A uniform rotation of the CMB polarization plane mixes the $E$- and $B$-mode polarizations and generates parity-odd $TB$ and $EB$ power spectra~\cite{Lue:1998mq,Feng:2004mq,Feng:2006dp}. For a global rotation angle $\beta$, the rotated CMB power spectra are related to the unrotated spectra by
\begin{align}
    C_\ell^{\prime TT} &= C_\ell^{TT},\\
    C_\ell^{\prime TE} &= C_\ell^{TE}\cos(2\beta),\\
    C_\ell^{\prime EE} &= C_\ell^{EE}\cos^2(2\beta)+C_\ell^{BB}\sin^2(2\beta),\\
    C_\ell^{\prime BB} &= C_\ell^{BB}\cos^2(2\beta)+C_\ell^{EE}\sin^2(2\beta),\\
    C_\ell^{\prime TB} &= C_\ell^{TE}\sin(2\beta),\\
    C_\ell^{\prime EB} &= \frac{1}{2}\left(C_\ell^{EE}-C_\ell^{BB}\right)\sin(4\beta)~.
\end{align}
These relations provide the basic principle for measuring a uniform CMB polarization rotation angle~\cite{Feng:2004mq,Feng:2006dp,Xia:2009ah,Li:2009rt}. The first measurement using WMAP and BOOMERANG data was performed in Ref.~\cite{Feng:2006dp}. Subsequent measurements and constraints were obtained by several CMB datasets, including QUaD, WMAP, ACTPol, SPTpol and Planck~\cite{QUaD:2008ado,WMAP:2012nax,ACTPol:2016kmo,Wu:2019hek,Planck:2016soo,Xia:2009ah,Xia:2012ck,Zhao:2015mqa}.

A central difficulty in measuring a global polarization rotation angle is its degeneracy with a global miscalibration of the detector polarization orientation. CMB experiments have used different calibration methods, including polarized astrophysical sources, artificial far-field sources, wire-grid calibration systems, optical modeling and Galactic foregrounds. The self-calibration method, which assumes the absence of physical polarization rotation and minimizes the observed $TB$ and $EB$ spectra, can reduce instrumental systematics but also removes any real global cosmic birefringence signal. To avoid this problem, recent analyses used Galactic foreground polarization or external calibration information to separate the instrumental miscalibration angle from the physical birefringence angle. Applying this strategy, Planck 2018 polarization data gave $\beta=0.35^\circ\pm0.14^\circ$, corresponding to a $2.4\sigma$ indication of global cosmic birefringence~\cite{Minami:2020odp}. A joint WMAP--Planck analysis improved this to $\beta=0.342^{\circ\,+0.094^\circ}_{\phantom{\circ\,}-0.091^\circ}$, excluding $\beta=0$ at $3.6\sigma$~\cite{Eskilt:2022cff}. More recently, ACT DR6 found $\beta=0.215^\circ\pm0.074^\circ$, a $2.9\sigma$ indication consistent with the WMAP and Planck results~\cite{Diego-Palazuelos:2025dmh}. These measurements are suggestive, although foreground modeling and instrumental systematics remain important issues.

In addition to the isotropic rotation, the polarization rotation angle may exhibit spatial variations, carrying information about fluctuations in the underlying dark-energy field that interacts with photons~\cite{Li:2008tma}. In this case, the rotation angle becomes a direction-dependent field $\beta(\hat{\mathbf n})$, and produces an anisotropic polarization-rotation pattern over the sky. The anisotropy can be statistically described by the angular power spectrum $C_L^{\beta\beta}$, analogously to the lensing-potential power spectrum $C_L^{\phi\phi}$. Assuming that the fluctuations of the external field obey statistical isotropy, non-perturbative expansion methods have been developed to relate the rotated and unrotated CMB power spectra~\cite{Li:2013vga,Zhao:2014yna}.

For a specific realization of the rotation pattern on the sky, however, statistical isotropy of the observed CMB is broken, leading to correlations between off-diagonal multipoles with $\ell\neq\ell'$. This mode coupling allows the rotation field to be reconstructed using quadratic-estimator techniques, closely analogous to CMB lensing reconstruction~\cite{Gluscevic:2009mm,Yadav:2009eb}. The first implementation of this method was carried out using WMAP7 data~\cite{Gluscevic:2012me}. Subsequent searches by the POLARBEAR, SPTpol, ACT, and BICEP/Keck collaborations have applied similar reconstruction methods, but so far no statistically significant anisotropic cosmic-birefringence signal has been detected~\cite{POLARBEAR:2015ktq,SPT:2020cxx,Namikawa:2020ffr,BICEPKeck:2022kci}. The best current 95\% upper bound on the amplitude of a scale-invariant rotation spectrum is
\begin{equation}
    A_{\rm CB}\leq 0.044,
\end{equation}
where $A_{\rm CB}$ is defined through
\begin{equation}
    \frac{L(L+1)C_L^{\beta\beta}}{2\pi}
    = A_{\rm CB}\times 10^{-4}\;{\rm rad}^2 .
\end{equation}

Recently,~\cite{Zhai:2025hqt} predicted the capability of the Ali CMB Polarization Telescope (AliCPT) to detect both isotropic and anisotropic CMB polarization rotation angles. For the isotropic case, the analysis jointly estimated the physical rotation angle and the instrumental polarization miscalibration angle, using Galactic foreground polarization and external calibration information to break their degeneracy. Taking the best-fit rotation angle inferred from Planck data as the fiducial input, it was found that AliCPT combined with Planck HFI data can reach a detection significance of about $5\sigma$ with 11 module-year observations, assuming a polarization-angle calibration precision of $0.1^\circ$. For anisotropic polarization rotation, the same work considered the sensitivity to the amplitude of a scale-invariant rotation spectrum. With a large-aperture AliCPT configuration and 50 module-year observations, the projected sensitivity can improve the current constraint significantly and provide a competitive test of spatial fluctuations in the dark energy field.

In summary, the Quintom dark energy model, as a very important category of dynamical dark energy, was proposed shortly after the discovery of cosmic accelerated expansion and has since been systematically studied. In the early 21st century, the observational constraints on the parameters of the CPL parameterization dark energy were not strict,  leaving the $\Lambda CDM$ model at the center of the allowed parameter space.  Following the DESI result, the current datasets show a preference in the parameter space of $w_0w_a$ for dynamical dark energy, especially the Quintom scenario. Given that parametrized $w(z)$ imposes limitations on the evolution of dark energy, non-parametric reconstruction methods for $w(z)$, such as Gaussian processes and principal component analysis, have also been developed. These model-independent approaches similarly support the Quintom dark energy scenario.

In addition to investigating the evolutionary features of dynamical dark energy from the perspective of observational data, it is also necessary to explore models capable of realizing Quintom dark energy from a theoretical standpoint. The No-Go theorem demonstrates that within the framework of general relativity, the EoS of a conventional single scalar field or a single fluid model cannot cross $w = -1$. Therefore, additional degrees of freedom are required to realize Quintom behavior. Since the first two-field Quintom model was proposed in 2004, the evolutionary characteristics of Quintom dark energy have been considered within various theoretical frameworks, including higher derivative term, modified gravity theories and interacting dark energy, highlighting the rich theoretical possibilities underlying the Quintom dark energy scenario.

Furthermore, the Quintom scenario provides some interesting behavior of the early universe. The bounce model presents a universe that starts with a contracting phase, then bounces into an expanding phase and evolves to the present day. The occurrence of this bounce requires the overall EoS of the universe to have Quintom-like behavior. Similarly, other non-singular cosmological models, such as the cyclic universe and the emergent universe, are also closely related to Quintom matter.

Dynamical dark energy may also interact with ordinary matter, for instance, through the Chern-Simons coupling to photons. Such an interaction would induce a rotation of the CMB polarization plane, making it potentially detectable by future CMB experiments.  This provides a new perspective for investigating the nature of dynamical dark energy.

{Forthcoming complete DESI dataset and new generation surveys like the Rubin Observatory's Legacy Survey of Space and Time~\cite{LSST:2008ijt}, Euclid~\cite{Euclid:2024yrr}, Simons Observatory~\cite{SimonsObservatory:2018koc}, AliCPT~\cite{Li:2017drr}, China Space Station Telescope~\cite{CSST:2025ssq} are expected to deliver unprecedented data with exceptional precision from  SN Ia, BAO, CMB, weak gravitational lensing and galaxy clustering. They will provide complementary and robust probes of the Quintom scenario. }

\section{ACKNOWLEDGEMENTS}
This work was supported in part by the National Key R$\&$D Program of China Grant (2021YFC2203100 and 2024YFC2207500), by the National Natural Science Foundation of China (92476203, 12247101), by the Fundamental Research Funds for the Central Universities (Grant No. lzujbky-2025-jdzx07), by the Natural Science Foundation of Gansu Province (No.25JRRA799), by the ‘111 Center’ under Grant No. B20063, by ``Talent Scientific Fund of Lanzhou University'' and CSC Innovation Talent Funds. XR thanks the support from the activity ``APCTP-2026-F02''.

\bibliography{Quintom}{}

@article{Xia:2005ge,
  author        = {Xia, Jun-Qing and Zhao, Gong-Bo and Feng, Bo and Li, Hong and Zhang, Xinmin},
  title         = {{Observing Dark Energy Dynamics with Supernova, Microwave Background and Galaxy Clustering}},
  journal       = {Phys. Rev. D},
  volume        = {73},
  pages         = {063521},
  year          = {2006},
  eprint        = {astro-ph/0511625},
  archivePrefix = {arXiv},
  doi           = {10.1103/PhysRevD.73.063521}
}

@article{Li:2010hm,
  author        = {Li, Mingzhe and Cai, Yi-Fu and Li, Hong and Brandenberger, Robert and Zhang, Xinmin},
  title         = {{Dark Energy Perturbations Revisited}},
  journal       = {Phys. Lett. B},
  volume        = {702},
  pages         = {5--11},
  year          = {2011},
  eprint        = {1008.1684},
  archivePrefix = {arXiv},
  primaryClass  = {astro-ph.CO},
  doi           = {10.1016/j.physletb.2011.06.042}
}

@article{Lue:1998mq,
    author = "Lue, Arthur and Wang, Li-Min and Kamionkowski, Marc",
    title = "{Cosmological signature of new parity violating interactions}",
    eprint = "astro-ph/9812088",
    archivePrefix = "arXiv",
    reportNumber = "CU-TP-926, CAL-675",
    doi = "10.1103/PhysRevLett.83.1506",
    journal = "Phys. Rev. Lett.",
    volume = "83",
    pages = "1506--1509",
    year = "1999"
}

@article{Feng:2004mq,
    author = "Feng, Bo and Li, Hong and Li, Mingzhe and Zhang, Xinmin",
    title = "{Gravitational leptogenesis and its signatures in CMB}",
    eprint = "hep-ph/0406269",
    archivePrefix = "arXiv",
    doi = "10.1016/j.physletb.2005.06.009",
    journal = "Phys. Lett. B",
    volume = "620",
    pages = "27--32",
    year = "2005"
}

@article{Xia:2009ah,
    author = "Xia, Jun-Qing and Li, Hong and Zhang, Xinmin",
    title = "{Probing CPT Violation with CMB Polarization Measurements}",
    eprint = "0908.1876",
    archivePrefix = "arXiv",
    primaryClass = "astro-ph.CO",
    doi = "10.1016/j.physletb.2010.03.038",
    journal = "Phys. Lett. B",
    volume = "687",
    pages = "129--132",
    year = "2010"
}

@article{Li:2009rt,
    author = "Li, Mingzhe and Cai, Yi-Fu and Wang, Xiulian and Zhang, Xinmin",
    title = "{$CPT$ Violating Electrodynamics and Chern-Simons Modified Gravity}",
    eprint = "0907.5159",
    archivePrefix = "arXiv",
    primaryClass = "hep-ph",
    doi = "10.1016/j.physletb.2009.08.053",
    journal = "Phys. Lett. B",
    volume = "680",
    pages = "118--124",
    year = "2009"
}

@article{QUaD:2008ado,
    author = "Wu, E. Y. S. and others",
    collaboration = "QUaD",
    title = "{Parity Violation Constraints Using Cosmic Microwave Background Polarization Spectra from 2006 and 2007 Observations by the QUaD Polarimeter}",
    eprint = "0811.0618",
    archivePrefix = "arXiv",
    primaryClass = "astro-ph",
    reportNumber = "SLAC-PUB-14791",
    doi = "10.1103/PhysRevLett.102.161302",
    journal = "Phys. Rev. Lett.",
    volume = "102",
    pages = "161302",
    year = "2009"
}

@article{WMAP:2012nax,
    author = "Hinshaw, G. and others",
    collaboration = "WMAP",
    title = "{Nine-Year Wilkinson Microwave Anisotropy Probe (WMAP) Observations: Cosmological Parameter Results}",
    eprint = "1212.5226",
    archivePrefix = "arXiv",
    primaryClass = "astro-ph.CO",
    doi = "10.1088/0067-0049/208/2/19",
    journal = "Astrophys. J. Suppl.",
    volume = "208",
    pages = "19",
    year = "2013"
}

@article{ACTPol:2016kmo,
    author = "Louis, Thibaut and others",
    collaboration = "ACTPol",
    title = "{The Atacama Cosmology Telescope: Two-Season ACTPol Spectra and Parameters}",
    eprint = "1610.02360",
    archivePrefix = "arXiv",
    primaryClass = "astro-ph.CO",
    doi = "10.1088/1475-7516/2017/06/031",
    journal = "JCAP",
    volume = "06",
    pages = "031",
    year = "2017"
}

@article{Planck:2016soo,
    author = "Aghanim, N. and others",
    collaboration = "Planck",
    title = "{Planck intermediate results. XLIX. Parity-violation constraints from polarization data}",
    eprint = "1605.08633",
    archivePrefix = "arXiv",
    primaryClass = "astro-ph.CO",
    doi = "10.1051/0004-6361/201629018",
    journal = "Astron. Astrophys.",
    volume = "596",
    pages = "A110",
    year = "2016"
}

@article{Wu:2019hek,
    author = "Wu, W. L. K. and others",
    title = "{A Measurement of the Cosmic Microwave Background Lensing Potential and Power Spectrum from 500 deg$^2$ of SPTpol Temperature and Polarization Data}",
    eprint = "1905.05777",
    archivePrefix = "arXiv",
    primaryClass = "astro-ph.CO",
    doi = "10.3847/1538-4357/ab4186",
    journal = "Astrophys. J.",
    volume = "884",
    pages = "70",
    year = "2019"
}

@article{Zhao:2015mqa,
    author = "Zhao, Gong-Bo and Wang, Yuting and Xia, Jun-Qing and Li, Mingzhe and Zhang, Xinmin",
    title = "{An efficient probe of the cosmological CPT violation}",
    eprint = "1504.04507",
    archivePrefix = "arXiv",
    primaryClass = "astro-ph.CO",
    doi = "10.1088/1475-7516/2015/07/032",
    journal = "JCAP",
    volume = "07",
    pages = "032",
    year = "2015"
}

@article{Xia:2012ck,
    author = "Xia, Jun-Qing",
    title = "{Cosmological CPT Violation and CMB Polarization Measurements}",
    eprint = "1201.4457",
    archivePrefix = "arXiv",
    primaryClass = "astro-ph.CO",
    doi = "10.1088/1475-7516/2012/01/046",
    journal = "JCAP",
    volume = "01",
    pages = "046",
    year = "2012"
}

@article{Minami:2020odp,
    author = "Minami, Yuto and Komatsu, Eiichiro",
    title = "{New Extraction of the Cosmic Birefringence from the Planck 2018 Polarization Data}",
    eprint = "2011.11254",
    archivePrefix = "arXiv",
    primaryClass = "astro-ph.CO",
    doi = "10.1103/PhysRevLett.125.221301",
    journal = "Phys. Rev. Lett.",
    volume = "125",
    number = "22",
    pages = "221301",
    year = "2020"
}

@article{Eskilt:2022cff,
    author = "Eskilt, Johannes R. and Komatsu, Eiichiro",
    title = "{Improved constraints on cosmic birefringence from the WMAP and Planck cosmic microwave background polarization data}",
    eprint = "2205.13962",
    archivePrefix = "arXiv",
    primaryClass = "astro-ph.CO",
    doi = "10.1103/PhysRevD.106.063503",
    journal = "Phys. Rev. D",
    volume = "106",
    number = "6",
    pages = "063503",
    year = "2022"
}

@ARTICLE{Diego-Palazuelos:2025dmh,
       author = {{Diego-Palazuelos}, P. and {Komatsu}, E.},
        title = "{Cosmic Birefringence from the Atacama Cosmology Telescope Data Release 6}",
      journal = {arXiv e-prints},
         year = 2025,
        month = "9",
          eid = {arXiv:2509.13654},
        pages = {arXiv:2509.13654},
          doi = {10.48550/arXiv.2509.13654},
archivePrefix = {arXiv},
       eprint = {2509.13654},
 primaryClass = {astro-ph.CO},
}

@article{Zhao:2014yna,
    author = "Zhao, Wen and Li, Mingzhe",
    title = "{Fluctuations of cosmological birefringence and the effect on CMB B-mode polarization}",
    eprint = "1403.3997",
    archivePrefix = "arXiv",
    primaryClass = "astro-ph.CO",
    doi = "10.1103/PhysRevD.89.103518",
    journal = "Phys. Rev. D",
    volume = "89",
    number = "10",
    pages = "103518",
    year = "2014"
}

@article{Gluscevic:2009mm,
    author = "Gluscevic, Vera and Kamionkowski, Marc and Cooray, Asantha",
    title = "{De-Rotation of the Cosmic Microwave Background Polarization: Full-Sky Formalism}",
    eprint = "0905.1687",
    archivePrefix = "arXiv",
    primaryClass = "astro-ph.CO",
    doi = "10.1103/PhysRevD.80.023510",
    journal = "Phys. Rev. D",
    volume = "80",
    pages = "023510",
    year = "2009"
}

@article{Yadav:2009eb,
    author = "Yadav, Amit P. S. and Biswas, Rahul and Su, Meng and Zaldarriaga, Matias",
    title = "{Constraining a spatially dependent rotation of the Cosmic Microwave Background Polarization}",
    eprint = "0902.4466",
    archivePrefix = "arXiv",
    primaryClass = "astro-ph.CO",
    doi = "10.1103/PhysRevD.79.123009",
    journal = "Phys. Rev. D",
    volume = "79",
    pages = "123009",
    year = "2009"
}

@article{Gluscevic:2012me,
  author = {Gluscevic, V. and Hanson, D. and Kamionkowski, M. and Hirata, C. M.},
  title = {First CMB Constraints on Direction-Dependent Cosmological Birefringence from WMAP-7},
  journal = {Phys. Rev. D},
  volume = {86},
  pages = {103529},
  year = {2012},
  doi = {10.1103/PhysRevD.86.103529},
  eprint = {1206.5546},
  archiveprefix = {arXiv},
  primaryclass = {astro-ph.CO}
}

@article{POLARBEAR:2015ktq,
    author = "Ade, Peter A. R. and others",
    collaboration = "POLARBEAR",
    title = "{POLARBEAR Constraints on Cosmic Birefringence and Primordial Magnetic Fields}",
    eprint = "1509.02461",
    archivePrefix = "arXiv",
    primaryClass = "astro-ph.CO",
    doi = "10.1103/PhysRevD.92.123509",
    journal = "Phys. Rev. D",
    volume = "92",
    pages = "123509",
    year = "2015"
}

@article{SPT:2020cxx,
    author = "Bianchini, F. and others",
    collaboration = "SPT",
    title = "{Searching for Anisotropic Cosmic Birefringence with Polarization Data from SPTpol}",
    eprint = "2006.08061",
    archivePrefix = "arXiv",
    primaryClass = "astro-ph.CO",
    reportNumber = "FERMILAB-PUB-20-264-AE",
    doi = "10.1103/PhysRevD.102.083504",
    journal = "Phys. Rev. D",
    volume = "102",
    number = "8",
    pages = "083504",
    year = "2020"
}

@article{Namikawa:2020ffr,
    author = "Namikawa, Toshiya and others",
    title = "{Atacama Cosmology Telescope: Constraints on cosmic birefringence}",
    eprint = "2001.10465",
    archivePrefix = "arXiv",
    primaryClass = "astro-ph.CO",
    doi = "10.1103/PhysRevD.101.083527",
    journal = "Phys. Rev. D",
    volume = "101",
    number = "8",
    pages = "083527",
    year = "2020"
}

@article{BICEPKeck:2022kci,
    author = "Ade, P. A. R. and others",
    collaboration = "BICEP/Keck",
    title = "{BICEP/Keck. XVII. Line-of-sight Distortion Analysis: Estimates of Gravitational Lensing, Anisotropic Cosmic Birefringence, Patchy Reionization, and Systematic Errors}",
    eprint = "2210.08038",
    archivePrefix = "arXiv",
    primaryClass = "astro-ph.CO",
    doi = "10.3847/1538-4357/acc85c",
    journal = "Astrophys. J.",
    volume = "949",
    number = "2",
    pages = "43",
    year = "2023"
}

@article{Cai:2012va,
    author = "Cai, Yi-Fu and Easson, Damien A. and Brandenberger, Robert",
    title = "{Towards a Nonsingular Bouncing Cosmology}",
    eprint = "1206.2382",
    archivePrefix = "arXiv",
    primaryClass = "hep-th",
    doi = "10.1088/1475-7516/2012/08/020",
    journal = "JCAP",
    volume = "08",
    pages = "020",
    year = "2012"
}

@article{Matsumoto:2017qil,
    author = "Matsumoto, Jiro",
    title = "{Phantom crossing dark energy in Horndeski\textquoteright{}s theory}",
    eprint = "1712.10015",
    archivePrefix = "arXiv",
    primaryClass = "gr-qc",
    doi = "10.1103/PhysRevD.97.123538",
    journal = "Phys. Rev. D",
    volume = "97",
    number = "12",
    pages = "123538",
    year = "2018"
}

@article{Tiwari:2024gzo,
    author = "Tiwari, Yashi and Upadhyay, Ujjwal and Jain, Rajeev Kumar",
    title = "{Exploring cosmological imprints of phantom crossing with dynamical dark energy in Horndeski gravity}",
    eprint = "2412.00931",
    archivePrefix = "arXiv",
    primaryClass = "astro-ph.CO",
    doi = "10.1103/PhysRevD.111.043530",
    journal = "Phys. Rev. D",
    volume = "111",
    number = "4",
    pages = "043530",
    year = "2025"
}

@article{Ellis:2002we,
    author = "Ellis, George F. R. and Maartens, Roy",
    title = "{The emergent universe: Inflationary cosmology with no singularity}",
    eprint = "gr-qc/0211082",
    archivePrefix = "arXiv",
    doi = "10.1088/0264-9381/21/1/015",
    journal = "Class. Quant. Grav.",
    volume = "21",
    pages = "223--232",
    year = "2004"
}

@article{Cai:2012yf,
    author = "Cai, Yi-Fu and Li, Mingzhe and Zhang, Xinmin",
    title = "{Emergent Universe Scenario via Quintom Matter}",
    eprint = "1209.3437",
    archivePrefix = "arXiv",
    primaryClass = "hep-th",
    doi = "10.1016/j.physletb.2012.10.065",
    journal = "Phys. Lett. B",
    volume = "718",
    pages = "248--254",
    year = "2012"
}

@article{Cai:2013rna,
    author = "Cai, Yi-Fu and Wan, Youping and Zhang, Xinmin",
    title = "{Cosmology of the Spinor Emergent Universe and Scale-invariant Perturbations}",
    eprint = "1312.0740",
    archivePrefix = "arXiv",
    primaryClass = "hep-th",
    doi = "10.1016/j.physletb.2014.02.042",
    journal = "Phys. Lett. B",
    volume = "731",
    pages = "217--226",
    year = "2014"
}

@article{Kolevatov:2017kvx,
    author = "Kolevatov, R. and Mironov, S. and Sukhov, N. and Volkova, V.",
    title = "{Cosmological bounce and Genesis beyond Horndeski}",
    eprint = "1705.06626",
    archivePrefix = "arXiv",
    primaryClass = "hep-th",
    reportNumber = "INR-TH-2017-015",
    doi = "10.1088/1475-7516/2017/08/038",
    journal = "JCAP",
    volume = "08",
    pages = "038",
    year = "2017"
}

@article{Xiong:2007cn,
    author = "Xiong, Hua-Hui and Qiu, Taotao and Cai, Yi-Fu and Zhang, Xinmin",
    title = "{Cyclic universe with quintom matter in loop quantum cosmology}",
    eprint = "0711.4469",
    archivePrefix = "arXiv",
    primaryClass = "hep-th",
    doi = "10.1142/S0217732309030667",
    journal = "Mod. Phys. Lett. A",
    volume = "24",
    pages = "1237--1246",
    year = "2009"
}

@article{SupernovaSearchTeam:1998fmf,
    author = "Riess, Adam G. and others",
    collaboration = "Supernova Search Team",
    title = "{Observational evidence from supernovae for an accelerating universe and a cosmological constant}",
    eprint = "astro-ph/9805201",
    archivePrefix = "arXiv",
    doi = "10.1086/300499",
    journal = "Astron. J.",
    volume = "116",
    pages = "1009--1038",
    year = "1998"
}

@article{SupernovaCosmologyProject:1998vns,
    author = "Perlmutter, S. and others",
    collaboration = "Supernova Cosmology Project",
    title = "{Measurements of $\Omega$ and $\Lambda$ from 42 High Redshift Supernovae}",
    eprint = "astro-ph/9812133",
    archivePrefix = "arXiv",
    reportNumber = "LBNL-41801, LBL-41801",
    doi = "10.1086/307221",
    journal = "Astrophys. J.",
    volume = "517",
    pages = "565--586",
    year = "1999"
}

@article{Feng:2004ad,
    author = "Feng, Bo and Wang, Xiu-Lian and Zhang, Xin-Min",
    title = "{Dark energy constraints from the cosmic age and supernova}",
    eprint = "astro-ph/0404224",
    archivePrefix = "arXiv",
    doi = "10.1016/j.physletb.2004.12.071",
    journal = "Phys. Lett. B",
    volume = "607",
    pages = "35--41",
    year = "2005"
}

@article{Cai:2015emx,
    author = "Cai, Yi-Fu and Capozziello, Salvatore and De Laurentis, Mariafelicia and Saridakis, Emmanuel N.",
    title = "{f(T) teleparallel gravity and cosmology}",
    eprint = "1511.07586",
    archivePrefix = "arXiv",
    primaryClass = "gr-qc",
    doi = "10.1088/0034-4885/79/10/106901",
    journal = "Rept. Prog. Phys.",
    volume = "79",
    number = "10",
    pages = "106901",
    year = "2016"
}

@article{BeltranJimenez:2019tme,
    author = "Beltr\'an Jim\'enez, Jose and Heisenberg, Lavinia and Koivisto, Tomi Sebastian and Pekar, Simon",
    title = "{Cosmology in $f(Q)$ geometry}",
    eprint = "1906.10027",
    archivePrefix = "arXiv",
    primaryClass = "gr-qc",
    doi = "10.1103/PhysRevD.101.103507",
    journal = "Phys. Rev. D",
    volume = "101",
    number = "10",
    pages = "103507",
    year = "2020"
}

@article{Heisenberg:2023lru,
    author = "Heisenberg, Lavinia",
    title = "{Review on f(Q) gravity}",
    eprint = "2309.15958",
    archivePrefix = "arXiv",
    primaryClass = "gr-qc",
    doi = "10.1016/j.physrep.2024.02.001",
    journal = "Phys. Rept.",
    volume = "1066",
    pages = "1--78",
    year = "2024"
}

@article{Basilakos:2025olm,
    author = "Basilakos, Spyros and Paliathanasis, Andronikos and Saridakis, Emmanuel N.",
    title = "{Equivalence of f(Q) cosmology with quintom-like scenario: The phantom field as effective realization of the non-trivial connection}",
    eprint = "2503.19864",
    archivePrefix = "arXiv",
    primaryClass = "gr-qc",
    doi = "10.1016/j.physletb.2025.139658",
    journal = "Phys. Lett. B",
    volume = "868",
    pages = "139658",
    year = "2025"
}

@article{Langlois:2017mxy,
    author = "Langlois, David and Mancarella, Michele and Noui, Karim and Vernizzi, Filippo",
    title = "{Effective Description of Higher-Order Scalar-Tensor Theories}",
    eprint = "1703.03797",
    archivePrefix = "arXiv",
    primaryClass = "hep-th",
    doi = "10.1088/1475-7516/2017/05/033",
    journal = "JCAP",
    volume = "05",
    pages = "033",
    year = "2017"
}

@article{Langlois:2018jdg,
    author = "Langlois, David and Mancarella, Michele and Noui, Karim and Vernizzi, Filippo",
    title = "{Mimetic gravity as DHOST theories}",
    eprint = "1802.03394",
    archivePrefix = "arXiv",
    primaryClass = "gr-qc",
    doi = "10.1088/1475-7516/2019/02/036",
    journal = "JCAP",
    volume = "02",
    pages = "036",
    year = "2019"
}

@article{Horndeski:1974wa,
    author = "Horndeski, Gregory Walter",
    title = "{Second-order scalar-tensor field equations in a four-dimensional space}",
    doi = "10.1007/BF01807638",
    journal = "Int. J. Theor. Phys.",
    volume = "10",
    pages = "363--384",
    year = "1974"
}

@article{DESI:2024mwx,
    author = "Adame, A. G. and others",
    collaboration = "DESI",
    title = "{DESI 2024 VI: cosmological constraints from the measurements of baryon acoustic oscillations}",
    eprint = "2404.03002",
    archivePrefix = "arXiv",
    primaryClass = "astro-ph.CO",
    reportNumber = "FERMILAB-PUB-24-0154-PPD",
    doi = "10.1088/1475-7516/2025/02/021",
    journal = "JCAP",
    volume = "02",
    pages = "021",
    year = "2025"
}

@article{DESI:2025zgx,
    author = "Abdul Karim, M. and others",
    collaboration = "DESI",
    title = "{DESI DR2 results. II. Measurements of baryon acoustic oscillations and cosmological constraints}",
    eprint = "2503.14738",
    archivePrefix = "arXiv",
    primaryClass = "astro-ph.CO",
    reportNumber = "FERMILAB-PUB-25-0169-PPD",
    doi = "10.1103/tr6y-kpc6",
    journal = "Phys. Rev. D",
    volume = "112",
    number = "8",
    pages = "083515",
    year = "2025"
}

@article{DESI:2025wyn,
    author = "Gu, Gan and others",
    collaboration = "DESI",
    title = "{Dynamical dark energy in light of the DESI DR2 baryonic acoustic oscillations measurements}",
    eprint = "2504.06118",
    archivePrefix = "arXiv",
    primaryClass = "astro-ph.CO",
    reportNumber = "FERMILAB-PUB-25-0235-PPD",
    doi = "10.1038/s41550-025-02669-6",
    journal = "Nature Astron.",
    volume = "9",
    number = "12",
    pages = "1879--1889",
    year = "2025",
    note = "[Erratum: Nature Astron. 9, 1898 (2025)]"
}

@article{Li:2011qfa,
    author = "Li, Mingzhe and Qiu, Taotao and Cai, Yifu and Zhang, Xinmin",
    title = "{On dark energy models of single scalar field}",
    eprint = "1112.4255",
    archivePrefix = "arXiv",
    primaryClass = "hep-th",
    doi = "10.1088/1475-7516/2012/04/003",
    journal = "JCAP",
    volume = "04",
    pages = "003",
    year = "2012"
}

@article{Cai:2009zp,
    author = "Cai, Yi-Fu and Saridakis, Emmanuel N. and Setare, Mohammad R. and Xia, Jun-Qing",
    title = "{Quintom Cosmology: Theoretical implications and observations}",
    eprint = "0909.2776",
    archivePrefix = "arXiv",
    primaryClass = "hep-th",
    doi = "10.1016/j.physrep.2010.04.001",
    journal = "Phys. Rept.",
    volume = "493",
    pages = "1--60",
    year = "2010"
}

@article{Yang:2025mws,
    author = "Yang, Yuhang and Wang, Qingqing and Ren, Xin and Saridakis, Emmanuel N. and Cai, Yi-Fu",
    title = "{Modified Gravity Realizations of Quintom Dark Energy after DESI DR2}",
    eprint = "2504.06784",
    archivePrefix = "arXiv",
    primaryClass = "astro-ph.CO",
    doi = "10.3847/1538-4357/ade43f",
    journal = "Astrophys. J.",
    volume = "988",
    number = "1",
    pages = "123",
    year = "2025"
}

@article{Vikman:2004dc,
    author = "Vikman, Alexander",
    title = "{Can dark energy evolve to the phantom?}",
    eprint = "astro-ph/0407107",
    archivePrefix = "arXiv",
    doi = "10.1103/PhysRevD.71.023515",
    journal = "Phys. Rev. D",
    volume = "71",
    pages = "023515",
    year = "2005"
}

@article{Chevallier:2000qy,
    author = "Chevallier, Michel and Polarski, David",
    title = "{Accelerating universes with scaling dark matter}",
    eprint = "gr-qc/0009008",
    archivePrefix = "arXiv",
    doi = "10.1142/S0218271801000822",
    journal = "Int. J. Mod. Phys. D",
    volume = "10",
    pages = "213--224",
    year = "2001"
}

@article{Linder:2002et,
    author = "Linder, Eric V.",
    title = "{Exploring the expansion history of the universe}",
    eprint = "astro-ph/0208512",
    archivePrefix = "arXiv",
    doi = "10.1103/PhysRevLett.90.091301",
    journal = "Phys. Rev. Lett.",
    volume = "90",
    pages = "091301",
    year = "2003"
}

@article{Shafieloo:2012ht,
    author = "Shafieloo, Arman and Kim, Alex G. and Linder, Eric V.",
    title = "{Gaussian Process Cosmography}",
    eprint = "1204.2272",
    archivePrefix = "arXiv",
    primaryClass = "astro-ph.CO",
    doi = "10.1103/PhysRevD.85.123530",
    journal = "Phys. Rev. D",
    volume = "85",
    pages = "123530",
    year = "2012"
}

@article{Seikel:2012uu,
    author = "Seikel, Marina and Clarkson, Chris and Smith, Mathew",
    title = "{Reconstruction of dark energy and expansion dynamics using Gaussian processes}",
    eprint = "1204.2832",
    archivePrefix = "arXiv",
    primaryClass = "astro-ph.CO",
    doi = "10.1088/1475-7516/2012/06/036",
    journal = "JCAP",
    volume = "06",
    pages = "036",
    year = "2012"
}

@article{Holsclaw:2010nb,
    author = "Holsclaw, Tracy and Alam, Ujjaini and Sanso, Bruno and Lee, Herbert and Heitmann, Katrin and Habib, Salman and Higdon, David",
    title = "{Nonparametric Reconstruction of the Dark Energy Equation of State}",
    eprint = "1009.5443",
    archivePrefix = "arXiv",
    primaryClass = "astro-ph.CO",
    reportNumber = "LA-UR-09-05888",
    doi = "10.1103/PhysRevD.82.103502",
    journal = "Phys. Rev. D",
    volume = "82",
    pages = "103502",
    year = "2010"
}

@article{Holsclaw:2010sk,
    author = "Holsclaw, Tracy and Alam, Ujjaini and Sanso, Bruno and Lee, Herbert and Heitmann, Katrin and Habib, Salman and Higdon, David",
    title = "{Nonparametric Dark Energy Reconstruction from Supernova Data}",
    eprint = "1011.3079",
    archivePrefix = "arXiv",
    primaryClass = "astro-ph.CO",
    reportNumber = "LA-UR-09-07764",
    doi = "10.1103/PhysRevLett.105.241302",
    journal = "Phys. Rev. Lett.",
    volume = "105",
    pages = "241302",
    year = "2010"
}

@article{DESI:2025fii,
    author = "Lodha, K. and others",
    collaboration = "DESI",
    title = "{Extended dark energy analysis using DESI DR2 BAO measurements}",
    eprint = "2503.14743",
    archivePrefix = "arXiv",
    primaryClass = "astro-ph.CO",
    reportNumber = "FERMILAB-PUB-25-0164-PPD",
    doi = "10.1103/w4c6-1r5j",
    journal = "Phys. Rev. D",
    volume = "112",
    number = "8",
    pages = "083511",
    year = "2025"
}

@article{Ma:1995ey,
    author = "Ma, Chung-Pei and Bertschinger, Edmund",
    title = "{Cosmological perturbation theory in the synchronous and conformal Newtonian gauges}",
    eprint = "astro-ph/9506072",
    archivePrefix = "arXiv",
    doi = "10.1086/176550",
    journal = "Astrophys. J.",
    volume = "455",
    pages = "7--25",
    year = "1995"
}

@article{Kodama:1984ziu,
    author = "Kodama, Hideo and Sasaki, Misao",
    title = "{Cosmological Perturbation Theory}",
    doi = "10.1143/PTPS.78.1",
    journal = "Prog. Theor. Phys. Suppl.",
    volume = "78",
    pages = "1--166",
    year = "1984"
}

@article{Bahamonde:2021gfp,
    author = "Bahamonde, Sebastian and Dialektopoulos, Konstantinos F. and Escamilla-Rivera, Celia and Farrugia, Gabriel and Gakis, Viktor and Hendry, Martin and Hohmann, Manuel and Levi Said, Jackson and Mifsud, Jurgen and Di Valentino, Eleonora",
    title = "{Teleparallel gravity: from theory to cosmology}",
    eprint = "2106.13793",
    archivePrefix = "arXiv",
    primaryClass = "gr-qc",
    doi = "10.1088/1361-6633/ac9cef",
    journal = "Rept. Prog. Phys.",
    volume = "86",
    number = "2",
    pages = "026901",
    year = "2023"
}

@article{Li:2005fm,
    author = "Li, Ming-zhe and Feng, Bo and Zhang, Xin-min",
    title = "{A Single scalar field model of dark energy with equation of state crossing -1}",
    eprint = "hep-ph/0503268",
    archivePrefix = "arXiv",
    doi = "10.1088/1475-7516/2005/12/002",
    journal = "JCAP",
    volume = "12",
    pages = "002",
    year = "2005"
}

@article{Zhao:2005vj,
    author = "Zhao, Gong-Bo and Xia, Jun-Qing and Li, Mingzhe and Feng, Bo and Zhang, Xinmin",
    title = "{Perturbations of the quintom models of dark energy and the effects on observations}",
    eprint = "astro-ph/0507482",
    archivePrefix = "arXiv",
    doi = "10.1103/PhysRevD.72.123515",
    journal = "Phys. Rev. D",
    volume = "72",
    pages = "123515",
    year = "2005"
}

@article{Huterer:2002hy,
    author = "Huterer, Dragan and Starkman, Glenn",
    title = "{Parameterization of dark-energy properties: A Principal-component approach}",
    eprint = "astro-ph/0207517",
    archivePrefix = "arXiv",
    reportNumber = "CWRU-07-02",
    doi = "10.1103/PhysRevLett.90.031301",
    journal = "Phys. Rev. Lett.",
    volume = "90",
    pages = "031301",
    year = "2003"
}

@article{Zhao:2012aw,
    author = "Zhao, Gong-Bo and Crittenden, Robert G. and Pogosian, Levon and Zhang, Xinmin",
    title = "{Examining the evidence for dynamical dark energy}",
    eprint = "1207.3804",
    archivePrefix = "arXiv",
    primaryClass = "astro-ph.CO",
    doi = "10.1103/PhysRevLett.109.171301",
    journal = "Phys. Rev. Lett.",
    volume = "109",
    pages = "171301",
    year = "2012"
}

@article{Zhao:2017cud,
    author = "Zhao, Gong-Bo and others",
    title = "{Dynamical dark energy in light of the latest observations}",
    eprint = "1701.08165",
    archivePrefix = "arXiv",
    primaryClass = "astro-ph.CO",
    doi = "10.1038/s41550-017-0216-z",
    journal = "Nature Astron.",
    volume = "1",
    number = "9",
    pages = "627--632",
    year = "2017"
}

@article{Starobinsky:1980te,
    author = "Starobinsky, Alexei A.",
    editor = "Khalatnikov, I. M. and Mineev, V. P.",
    title = "{A New Type of Isotropic Cosmological Models Without Singularity}",
    doi = "10.1016/0370-2693(80)90670-X",
    journal = "Phys. Lett. B",
    volume = "91",
    pages = "99--102",
    year = "1980"
}

@article{Cai:2007qw,
    author = "Cai, Yi-Fu and Qiu, Taotao and Piao, Yun-Song and Li, Mingzhe and Zhang, Xinmin",
    title = "{Bouncing universe with quintom matter}",
    eprint = "0704.1090",
    archivePrefix = "arXiv",
    primaryClass = "gr-qc",
    doi = "10.1088/1126-6708/2007/10/071",
    journal = "JHEP",
    volume = "10",
    pages = "071",
    year = "2007"
}

@article{Wei:2007rp,
    author = "Wei, Hao and Zhang, Shuang Nan",
    title = "{Dynamics of Quintom and Hessence Energies in Loop Quantum Cosmology}",
    eprint = "0705.4002",
    archivePrefix = "arXiv",
    primaryClass = "gr-qc",
    doi = "10.1103/PhysRevD.76.063005",
    journal = "Phys. Rev. D",
    volume = "76",
    pages = "063005",
    year = "2007"
}

@article{Cai:2008qb,
    author = "Cai, Yi-Fu and Qiu, Taotao and Xia, Jun-Qing and Li, Hong and Zhang, Xinmin",
    title = "{A Model Of Inflationary Cosmology Without Singularity}",
    eprint = "0808.0819",
    archivePrefix = "arXiv",
    primaryClass = "astro-ph",
    doi = "10.1103/PhysRevD.79.021303",
    journal = "Phys. Rev. D",
    volume = "79",
    pages = "021303",
    year = "2009"
}

@article{Cai:2008qw,
    author = "Cai, Yi-Fu and Qiu, Tao-tao and Brandenberger, Robert and Zhang, Xin-min",
    title = "{A Nonsingular Cosmology with a Scale-Invariant Spectrum of Cosmological Perturbations from Lee-Wick Theory}",
    eprint = "0810.4677",
    archivePrefix = "arXiv",
    primaryClass = "hep-th",
    doi = "10.1103/PhysRevD.80.023511",
    journal = "Phys. Rev. D",
    volume = "80",
    pages = "023511",
    year = "2009"
}

@article{Cai:2007zv,
    author = "Cai, Yi-Fu and Qiu, Taotao and Brandenberger, Robert and Piao, Yun-Song and Zhang, Xinmin",
    title = "{On Perturbations of Quintom Bounce}",
    eprint = "0711.2187",
    archivePrefix = "arXiv",
    primaryClass = "hep-th",
    reportNumber = "CAS-KITPC-ITP-016",
    doi = "10.1088/1475-7516/2008/03/013",
    journal = "JCAP",
    volume = "03",
    pages = "013",
    year = "2008"
}

@article{Cai:2008ed,
    author = "Cai, Yi-Fu and Zhang, Xinmin",
    title = "{Evolution of Metric Perturbations in Quintom Bounce model}",
    eprint = "0808.2551",
    archivePrefix = "arXiv",
    primaryClass = "astro-ph",
    doi = "10.1088/1475-7516/2009/06/003",
    journal = "JCAP",
    volume = "06",
    pages = "003",
    year = "2009"
}

@article{Cai:2009fn,
    author = "Cai, Yi-Fu and Xue, Wei and Brandenberger, Robert and Zhang, Xinmin",
    title = "{Non-Gaussianity in a Matter Bounce}",
    eprint = "0903.0631",
    archivePrefix = "arXiv",
    primaryClass = "astro-ph.CO",
    doi = "10.1088/1475-7516/2009/05/011",
    journal = "JCAP",
    volume = "05",
    pages = "011",
    year = "2009"
}

@article{Steinhardt:2002ih,
    author = "Steinhardt, Paul J. and Turok, Neil and Turok, N.",
    title = "{A Cyclic model of the universe}",
    eprint = "hep-th/0111030",
    archivePrefix = "arXiv",
    doi = "10.1126/science.1070462",
    journal = "Science",
    volume = "296",
    pages = "1436--1439",
    year = "2002"
}

@article{Ellis:2003qz,
    author = "Ellis, George F. R. and Murugan, Jeff and Tsagas, Christos G.",
    title = "{The Emergent universe: An Explicit construction}",
    eprint = "gr-qc/0307112",
    archivePrefix = "arXiv",
    doi = "10.1088/0264-9381/21/1/016",
    journal = "Class. Quant. Grav.",
    volume = "21",
    number = "1",
    pages = "233--250",
    year = "2004"
}

@article{Capozziello:2002rd,
    author = "Capozziello, Salvatore",
    title = "{Curvature quintessence}",
    eprint = "gr-qc/0201033",
    archivePrefix = "arXiv",
    doi = "10.1142/S0218271802002025",
    journal = "Int. J. Mod. Phys. D",
    volume = "11",
    pages = "483--492",
    year = "2002"
}

@article{Krssak:2018ywd,
    author = {Krssak, M. and van den Hoogen, R.J. and Pereira, J.G. and B\"ohmer, C.G. and Coley, A.A.},
    title = "{Teleparallel theories of gravity: illuminating a fully invariant approach}",
    eprint = "1810.12932",
    archivePrefix = "arXiv",
    primaryClass = "gr-qc",
    doi = "10.1088/1361-6382/ab2e1f",
    journal = "Class. Quant. Grav.",
    volume = "36",
    number = "18",
    pages = "183001",
    year = "2019"
}

@article{Nojiri:2010wj,
    author = "Nojiri, Shin'ichi and Odintsov, Sergei D.",
    title = "{Unified cosmic history in modified gravity: from F(R) theory to Lorentz non-invariant models}",
    eprint = "1011.0544",
    archivePrefix = "arXiv",
    primaryClass = "gr-qc",
    doi = "10.1016/j.physrep.2011.04.001",
    journal = "Phys. Rept.",
    volume = "505",
    pages = "59--144",
    year = "2011"
}

@article{Ren:2021tfi,
    author = "Ren, Xin and Wong, Thomas Hong Tsun and Cai, Yi-Fu and Saridakis, Emmanuel N.",
    title = "{Data-driven Reconstruction of the Late-time Cosmic Acceleration with f(T) Gravity}",
    eprint = "2103.01260",
    archivePrefix = "arXiv",
    primaryClass = "astro-ph.CO",
    doi = "10.1016/j.dark.2021.100812",
    journal = "Phys. Dark Univ.",
    volume = "32",
    pages = "100812",
    year = "2021"
}

@article{Yang:2025kgc,
    author = "Yang, Yuhang and Wang, Qingqing and Li, Chunyu and Yuan, Peibo and Ren, Xin and Saridakis, Emmanuel N. and Cai, Yi-Fu",
    title = "{Gaussian process reconstructions and model building of quintom dark energy from latest cosmological observations}",
    eprint = "2501.18336",
    archivePrefix = "arXiv",
    primaryClass = "astro-ph.CO",
    doi = "10.1088/1475-7516/2025/08/050",
    journal = "JCAP",
    volume = "08",
    pages = "050",
    year = "2025"
}

@article{Briffa:2020qli,
    author = "Briffa, Rebecca and Capozziello, Salvatore and Levi Said, Jackson and Mifsud, Jurgen and Saridakis, Emmanuel N.",
    title = "{Constraining teleparallel gravity through Gaussian processes}",
    eprint = "2009.14582",
    archivePrefix = "arXiv",
    primaryClass = "gr-qc",
    doi = "10.1088/1361-6382/abd4f5",
    journal = "Class. Quant. Grav.",
    volume = "38",
    number = "5",
    pages = "055007",
    year = "2020"
}

@article{Li:2008tma,
    author = "Li, Mingzhe and Zhang, Xinmin",
    title = "{Cosmological CPT violating effect on CMB polarization}",
    eprint = "0810.0403",
    archivePrefix = "arXiv",
    primaryClass = "astro-ph",
    doi = "10.1103/PhysRevD.78.103516",
    journal = "Phys. Rev. D",
    volume = "78",
    pages = "103516",
    year = "2008"
}

@article{Feng:2006dp,
    author = "Feng, Bo and Li, Mingzhe and Xia, Jun-Qing and Chen, Xuelei and Zhang, Xinmin",
    title = "{Searching for CPT Violation with Cosmic Microwave Background Data from WMAP and BOOMERANG}",
    eprint = "astro-ph/0601095",
    archivePrefix = "arXiv",
    reportNumber = "RESCEU-2-06",
    doi = "10.1103/PhysRevLett.96.221302",
    journal = "Phys. Rev. Lett.",
    volume = "96",
    pages = "221302",
    year = "2006"
}

@article{Li:2013vga,
    author = "Li, Mingzhe and Yu, Bo",
    title = "{New Constraints on Anisotropic Rotation of CMB Polarization}",
    eprint = "1303.1881",
    archivePrefix = "arXiv",
    primaryClass = "astro-ph.CO",
    reportNumber = "USTC-ICTS-13-03",
    doi = "10.1088/1475-7516/2013/06/016",
    journal = "JCAP",
    volume = "06",
    pages = "016",
    year = "2013"
}

@article{Li:2017drr,
    author = "Li, Hong and others",
    title = "{Probing Primordial Gravitational Waves: Ali CMB Polarization Telescope}",
    eprint = "1710.03047",
    archivePrefix = "arXiv",
    primaryClass = "astro-ph.CO",
    doi = "10.1093/nsr/nwy019",
    journal = "Natl. Sci. Rev.",
    volume = "6",
    number = "1",
    pages = "145--154",
    year = "2019"
}

@article{Zhang:2005eg,
    author = "Zhang, Xiao-Fei and Li, Hong and Piao, Yun-Song and Zhang, Xin-Min",
    title = "{Two-field models of dark energy with equation of state across -1}",
    eprint = "astro-ph/0501652",
    archivePrefix = "arXiv",
    doi = "10.1142/S0217732306018469",
    journal = "Mod. Phys. Lett. A",
    volume = "21",
    pages = "231--242",
    year = "2006"
}

@article{Khoury:2025txd,
    author = "Khoury, Justin and Lin, Meng-Xiang and Trodden, Mark",
    title = "{Apparent w{\ensuremath{<}}-1 and a Lower S8 from Dark Axion and Dark Baryons Interactions}",
    eprint = "2503.16415",
    archivePrefix = "arXiv",
    primaryClass = "astro-ph.CO",
    doi = "10.1103/w4qb-plk8",
    journal = "Phys. Rev. Lett.",
    volume = "135",
    number = "18",
    pages = "181001",
    year = "2025"
}

@article{Zhao:2009ke,
    author = "Zhao, Hongsheng and Macci\`o, Andrea V. and Li, Baojiu and Hoekstra, Henk and Feix, Martin",
    title = "{Structure Formation by Fifth Force: Power Spectrum from N-Body Simulations}",
    eprint = "0910.3207",
    archivePrefix = "arXiv",
    primaryClass = "astro-ph.CO",
    doi = "10.1088/2041-8205/712/2/L179",
    journal = "Astrophys. J. Lett.",
    volume = "712",
    pages = "L179--L183",
    year = "2010"
}

@article{Wang:2005jx,
    author = "Wang, Bin and Gong, Yun-gui and Abdalla, Elcio",
    title = "{Transition of the dark energy equation of state in an interacting holographic dark energy model}",
    eprint = "hep-th/0506069",
    archivePrefix = "arXiv",
    doi = "10.1016/j.physletb.2005.08.008",
    journal = "Phys. Lett. B",
    volume = "624",
    pages = "141--146",
    year = "2005"
}

@article{Wang:2005ph,
    author = "Wang, Bin and Lin, Chi-Yong and Abdalla, Elcio",
    title = "{Constraints on the interacting holographic dark energy model}",
    eprint = "hep-th/0509107",
    archivePrefix = "arXiv",
    doi = "10.1016/j.physletb.2006.04.009",
    journal = "Phys. Lett. B",
    volume = "637",
    pages = "357--361",
    year = "2006"
}

@article{Cortes:2024lgw,
    author = "Cort\^es, Marina and Liddle, Andrew R.",
    title = "{Interpreting DESI's evidence for evolving dark energy}",
    eprint = "2404.08056",
    archivePrefix = "arXiv",
    primaryClass = "astro-ph.CO",
    doi = "10.1088/1475-7516/2024/12/007",
    journal = "JCAP",
    volume = "12",
    pages = "007",
    year = "2024"
}

@article{Hu:2004kh,
    author = "Hu, Wayne",
    title = "{Crossing the phantom divide: Dark energy internal degrees of freedom}",
    eprint = "astro-ph/0410680",
    archivePrefix = "arXiv",
    doi = "10.1103/PhysRevD.71.047301",
    journal = "Phys. Rev. D",
    volume = "71",
    pages = "047301",
    year = "2005"
}

@article{DESI:2024aqx,
    author = "Calderon, R. and others",
    collaboration = "DESI",
    title = "{DESI 2024: reconstructing dark energy using crossing statistics with DESI DR1 BAO data}",
    eprint = "2405.04216",
    archivePrefix = "arXiv",
    primaryClass = "astro-ph.CO",
    doi = "10.1088/1475-7516/2024/10/048",
    journal = "JCAP",
    volume = "10",
    pages = "048",
    year = "2024"
}

@article{DESI:2024kob,
    author = "Lodha, K. and others",
    collaboration = "DESI",
    title = "{DESI 2024: Constraints on physics-focused aspects of dark energy using DESI DR1 BAO data}",
    eprint = "2405.13588",
    archivePrefix = "arXiv",
    primaryClass = "astro-ph.CO",
    reportNumber = "FERMILAB-PUB-24-0756-PPD",
    doi = "10.1103/PhysRevD.111.023532",
    journal = "Phys. Rev. D",
    volume = "111",
    number = "2",
    pages = "023532",
    year = "2025"
}

@article{Wang:2024dka,
    author = "Wang, Hao and Piao, Yun-Song",
    title = "{Dark energy in light of DESI DR1 and Hubble tension}",
    eprint = "2404.18579",
    archivePrefix = "arXiv",
    primaryClass = "astro-ph.CO",
    doi = "10.1016/j.physletb.2026.140180",
    journal = "Phys. Lett. B",
    volume = "873",
    pages = "140180",
    year = "2026"
}

@article{Giare:2024gpk,
    author = "Giar\`e, William and Najafi, Mahdi and Pan, Supriya and Di Valentino, Eleonora and Firouzjaee, Javad T.",
    title = "{Robust preference for Dynamical Dark Energy in DESI BAO and SN measurements}",
    eprint = "2407.16689",
    archivePrefix = "arXiv",
    primaryClass = "astro-ph.CO",
    doi = "10.1088/1475-7516/2024/10/035",
    journal = "JCAP",
    volume = "10",
    pages = "035",
    year = "2024"
}

@article{Mukherjee:2024ryz,
    author = "Mukherjee, Purba and Sen, Anjan Ananda",
    title = "{Model-independent cosmological inference post DESI DR1 BAO measurements}",
    eprint = "2405.19178",
    archivePrefix = "arXiv",
    primaryClass = "astro-ph.CO",
    doi = "10.1103/PhysRevD.110.123502",
    journal = "Phys. Rev. D",
    volume = "110",
    number = "12",
    pages = "123502",
    year = "2024"
}

@article{Jiang:2024xnu,
    author = "Jiang, Jun-Qian and Pedrotti, Davide and da Costa, Simony Santos and Vagnozzi, Sunny",
    title = "{Nonparametric late-time expansion history reconstruction and implications for the Hubble tension in light of recent DESI and type Ia supernovae data}",
    eprint = "2408.02365",
    archivePrefix = "arXiv",
    primaryClass = "astro-ph.CO",
    doi = "10.1103/PhysRevD.110.123519",
    journal = "Phys. Rev. D",
    volume = "110",
    number = "12",
    pages = "123519",
    year = "2024"
}

@article{Dinda:2024kjf,
    author = "Dinda, Bikash R.",
    title = "{A new diagnostic for the null test of dynamical dark energy in light of DESI 2024 and other BAO data}",
    eprint = "2405.06618",
    archivePrefix = "arXiv",
    primaryClass = "astro-ph.CO",
    doi = "10.1088/1475-7516/2024/09/062",
    journal = "JCAP",
    volume = "09",
    pages = "062",
    year = "2024"
}

@article{Giare:2024oil,
    author = "Giar{\`e}, William",
    title = "{Dynamical dark energy beyond Planck? Constraints from multiple CMB probes, DESI BAO, and type-Ia supernovae}",
    eprint = "2409.17074",
    archivePrefix = "arXiv",
    primaryClass = "astro-ph.CO",
    doi = "10.1103/ss37-cxhn",
    journal = "Phys. Rev. D",
    volume = "112",
    number = "2",
    pages = "023508",
    year = "2025"
}

@article{Carloni:2024zpl,
    author = "Carloni, Youri and Luongo, Orlando and Muccino, Marco",
    title = "{Does dark energy really revive using DESI 2024 data?}",
    eprint = "2404.12068",
    archivePrefix = "arXiv",
    primaryClass = "astro-ph.CO",
    doi = "10.1103/PhysRevD.111.023512",
    journal = "Phys. Rev. D",
    volume = "111",
    number = "2",
    pages = "023512",
    year = "2025"
}

@article{Gialamas:2024lyw,
    author = {Gialamas, Ioannis D. and H{\"u}tsi, Gert and Kannike, Kristjan and Racioppi, Antonio and Raidal, Martti and Vasar, Martin and Veerm{\"a}e, Hardi},
    title = "{Interpreting DESI 2024 BAO: Late-time dynamical dark energy or a local effect?}",
    eprint = "2406.07533",
    archivePrefix = "arXiv",
    primaryClass = "astro-ph.CO",
    doi = "10.1103/PhysRevD.111.043540",
    journal = "Phys. Rev. D",
    volume = "111",
    number = "4",
    pages = "043540",
    year = "2025"
}

@article{Liu:2024gfy,
    author = "Liu, Guanlin and Wang, Yu and Zhao, Wen",
    title = "{Impact of LRG1 and LRG2 in DESI 2024 BAO data on dark energy evolution}",
    eprint = "2407.04385",
    archivePrefix = "arXiv",
    primaryClass = "astro-ph.CO",
    month = "7",
    year = "2024"
}

@article{Escamilla-Rivera:2024sae,
    author = "Escamilla-Rivera, Celia and Sandoval-Orozco, Rodrigo",
    title = "{f(T) gravity after DESI Baryon acoustic oscillation and DES supernovae 2024 data}",
    eprint = "2405.00608",
    archivePrefix = "arXiv",
    primaryClass = "astro-ph.CO",
    doi = "10.1016/j.jheap.2024.05.005",
    journal = "JHEAp",
    volume = "42",
    pages = "217--221",
    year = "2024"
}

@article{Yin:2024hba,
    author = "Yin, Wen",
    title = "{Cosmic clues: DESI, dark energy, and the cosmological constant problem}",
    eprint = "2404.06444",
    archivePrefix = "arXiv",
    primaryClass = "hep-ph",
    doi = "10.1007/JHEP05(2024)327",
    journal = "JHEP",
    volume = "05",
    pages = "327",
    year = "2024"
}

@article{Chudaykin:2024gol,
    author = "Chudaykin, Anton and Kunz, Martin",
    title = "{Modified gravity interpretation of the evolving dark energy in light of DESI data}",
    eprint = "2407.02558",
    archivePrefix = "arXiv",
    primaryClass = "astro-ph.CO",
    doi = "10.1103/PhysRevD.110.123524",
    journal = "Phys. Rev. D",
    volume = "110",
    number = "12",
    pages = "123524",
    year = "2024"
}

@article{Huang:2025som,
    author = "Huang, Lu and Cai, Rong-Gen and Wang, Shao-Jiang",
    title = "{The DESI DR1/DR2 evidence for dynamical dark energy is biased by low-redshift supernovae}",
    eprint = "2502.04212",
    archivePrefix = "arXiv",
    primaryClass = "astro-ph.CO",
    doi = "10.1007/s11433-025-2754-5",
    journal = "Sci. China Phys. Mech. Astron.",
    volume = "68",
    number = "10",
    pages = "100413",
    year = "2025"
}

@article{RoyChoudhury:2024wri,
    author = "Roy Choudhury, Shouvik and Okumura, Teppei",
    title = "{Updated Cosmological Constraints in Extended Parameter Space with Planck PR4, DESI Baryon Acoustic Oscillations, and Supernovae: Dynamical Dark Energy, Neutrino Masses, Lensing Anomaly, and the Hubble Tension}",
    eprint = "2409.13022",
    archivePrefix = "arXiv",
    primaryClass = "astro-ph.CO",
    doi = "10.3847/2041-8213/ad8c26",
    journal = "Astrophys. J. Lett.",
    volume = "976",
    number = "1",
    pages = "L11",
    year = "2024"
}
\bibliographystyle{elsarticle-num}

\end{document}